\global\def\draftcontrol{0}

   \def\versionno{ABCD of Beta Ensembles}

\catcode`\@=11

\expandafter\ifx\csname draftcontrol\endcsname\relax\global\def\draftcontrol{0} 
\fi 

{\count255=\time\divide\count255 by 60 
\xdef\hourmin{\number\count255} 
\multiply\count255 by-60\advance\count255 by\time 
\xdef\hourmin{\hourmin:\ifnum\count255<10 0\fi\the\count255}} 
\def\draftdate{\number\month/\number\day/\number\year\ \ \ \hourmin } 

\newcommand\makepapertitle{\par

  \begingroup 
    \renewcommand\thefootnote{\@fnsymbol\c@footnote}%
    \def\@makefnmark{\rlap{\@textsuperscript{\normalfont\@thefnmark}}}%
    \long\def\@makefntext##1{\parindent 1em\noindent 
            \hb@xt@1.8em{%
                \hss\@textsuperscript{\normalfont\@thefnmark}}##1}%
     \newpage 
     \global\@topnum\z@   
     \@makepapertitle 
     \thispagestyle{empty}\@thanks 
  \endgroup 
  \setcounter{footnote}{0}%
  \global\let\thanks\relax 
  \global\let\makepapertitle\relax 
  \global\let\@makepapertitle\relax 
  \global\let\@thanks\@empty 
  \global\let\@author\@empty 
  \global\let\@date\@empty 
  \global\let\@title\@empty 
  \global\let\title\relax 
  \global\let\author\relax 
  \global\let\date\relax 
  \global\let\and\relax 
  \def\version{\let\version\@version\@gobble} 
} 
\def\@makepapertitle{%
  \newpage 
   \ifnum\draftcontrol=1 {} 
   \version\versionno 
   \vskip 5.5em%
   \else 
   \hfill\hbox to 3cm {\parbox{4.5cm}{\@pubnum}\hss}%
   \vskip 6.5em%
   \fi 
   \begin{center}%
   \let \footnote \thanks 
      {\hskip -0\textwidth \hbox to 1\textwidth%
        {\centerline{\Large\bf{\noindent\@title}}}}%
     \vskip 2em%
     {\normalsize
       \lineskip .5em%
       \begin{tabular}[t]{c}%
         \@author 
       \end{tabular}\par}%
     \vskip 1.5em%
     {\@bstract}%
     \end{center}%
     \vfill
     \@date%
     \vskip 1.5em%
   \par 
} 

\gdef\@pubnum{} 
\def\pubnum#1{%
  \gdef\@pubnum{#1}} 

\gdef\@bstract{} 
\def\Abstract#1{%
  \gdef\@bstract{%
   \parbox{\textwidth-0pc}{%
   \centerline{\bf Abstract}\penalty1000 
   \noindent
   \renewcommand\baselinestretch{1.0} 
   {#1}}} 
} 

\gdef\@email{}
\def\email#1{%
   \gdef\@email{%
   Email: {\tt #1}}
}

\def\ps@paper{\let\@mkboth\@gobbletwo%
     \ifnum\draftcontrol=1 
        \def\@oddfoot{\hbox to \textwidth{\tiny \versionno \hfil\tiny\draftdate}%
        \hskip -\textwidth \hbox to \textwidth{\hfil\rm\thepage\hfil}}%
     \else\def\@oddfoot{\hbox to \textwidth{\hfil\rm\thepage\hfil}} 
     \fi 
     \let\@evenfoot\@oddfoot 
} 

\def\body{\clearpage 
          \pagestyle{paper} 
        } 
\newenvironment{acknowledgments}{%
\vskip 3.25ex 
\addcontentsline{toc}{section}{Acknowledgments}
\noindent {\bf Acknowledgments} 
} 

\def\@version#1{\ifnum\draftcontrol=1 
\typeout{}\typeout{#1}\typeout{} 
\vskip3mm\centerline{\hbox{\fbox{\normalsize{\tt DRAFT -- #1 -- } 
                   {\draftdate}}}}\vskip3mm 
\fi} 
\let\version\@version 
\long\def\eqlabel#1{\ifnum\draftcontrol=1 
                    \tag@false  
                    \tag*{(\theequation) \hbox to -0.2cm{\hspace{0cm}\small{#1}\hss}} 
                    \refstepcounter{equation}  
                    \edef\@currentlabel{\theequation} 
                    \ltx@label{#1}          
                    \else 
                    \label{#1} 
                    \fi 
                    } 
\let\st@bibitem\@bibitem 
\let\st@lbibitem\@lbibitem 
\ifnum\draftcontrol=1 
  \def\@bibitem#1{%
    \st@bibitem{#1}\a@@label{#1}\ignorespaces} 
  \def\@lbibitem[#1]#2{%
    \st@lbibitem[#1]{#2}\a@@label{#2}\ignorespaces} 
  \def\a@@label#1{%
    \gdef\a@lab{\smash{\normalfont\small#1}} 
    \ifvmode 
      \if@inlabel 
        \global\setbox\@labels\hbox{%
          \llap{\a@lab\let\a@lab\relax 
                \kern\@totalleftmargin\kern\marginparsep}%
          \box\@labels}%
      \fi 
    \fi} 
\fi 

\documentclass[12pt,letterpaper]{article} 

\usepackage{amsmath,bm,amsfonts,amssymb,array,calc,amsthm,rotating}
\usepackage{epsfig,psfrag} 
\usepackage{rotating}
\usepackage{amscd}
\usepackage{graphicx}
\usepackage{color}
\usepackage[colorlinks=false]{hyperref}

\renewcommand\baselinestretch{1.25} 
\renewcommand\section{\@startsection {section}{1}{\z@}%
                                   {-3.5ex \@plus -1ex \@minus -.2ex}%
                                   {2.3ex \@plus.2ex}%
                                   {\normalfont\large\bfseries}} 
\renewcommand\subsection{\@startsection{subsection}{2}{\z@}%
                                   {-3.25ex\@plus -1ex \@minus -.2ex}%
                                   {1.5ex \@plus .2ex}%
                                   {\normalfont\normalsize\bfseries}} 
\renewcommand\subsubsection{\@startsection{subsubsection}{3}{\z@}%
                                   {-3.25ex\@plus -1ex \@minus -.2ex}%
                                   {1.5ex \@plus .2ex}%
                                   {\normalfont\normalsize\it}} 
\renewcommand\paragraph{\@startsection{paragraph}{4}{\z@}%
                                   {-1.75ex\@plus -1ex \@minus -.2ex}%
                                   {1ex \@plus .2ex}%
                                   {\normalfont\normalsize\bf}} 
\renewcommand\subparagraph{\@startsection{subparagraph}{5}{\z@}%
                                   {-1.25ex\@plus -0ex \@minus -.2ex}%
                                   {-2ex \@plus .2ex}%
                                   {\normalfont\normalsize\it}}

\numberwithin{equation}{section}

\long\def\@makecaption#1#2{%
  \vskip\abovecaptionskip
  \sbox\@tempboxa{{\bf #1:} #2}%
  \ifdim \wd\@tempboxa >\hsize
    {\small\bf #1:} {\small #2}\par
  \else
    \global \@minipagefalse
    \hb@xt@\hsize{\hfil\box\@tempboxa\hfil}%
  \fi
  \vskip\belowcaptionskip}

\renewcommand*\l@section[2]{%
  \ifnum \c@tocdepth >\z@
    \addpenalty\@secpenalty
    \addvspace{.5em \@plus\p@}%
    \setlength\@tempdima{1.5em}%
    \begingroup
      \parindent \z@ \rightskip \@pnumwidth
      \parfillskip -\@pnumwidth
      \leavevmode \bfseries
      \advance\leftskip\@tempdima
      \hskip -\leftskip
      #1\nobreak\hfil \nobreak\hb@xt@\@pnumwidth{\hss #2}\par
    \endgroup
  \fi}
\renewcommand*\l@subsection{\addvspace{.0em \@plus\p@}\@dottedtocline{2}{1.5em}{2.3em}}
\renewcommand*\l@subsubsection{\addvspace{-.2em \@plus\p@}\@dottedtocline{3}{3.8em}{3.2em}}

\def\hepth#1{\href{http://xxx.arxiv.org/abs/hep-th/#1}{{arXiv:hep-th/#1}}}

\def\arxiv#1#2{\href{http://xxx.arxiv.org/abs/#1}{{arXiv:#1 [#2]}}}

\definecolor{refcol}{rgb}{0.2,0.2,0.8}
\definecolor{eqcol}{rgb}{.6,0,0}
\definecolor{purple}{cmyk}{0,1,0,0}

\gdef\@citecolor{refcol}
\gdef\@linkcolor{eqcol}
\def\colorlinkspurple{\gdef\@urlcolor{purple}}
\def\colorlinksblue{\gdef\@urlcolor{blue}}
\def\colorlinksred{\gdef\@urlcolor{red}}

\def\ie{{\it i.e.}}

\def\cf{{\it cf.}}

\def\revise#1       {\raisebox{-0em}{\rule{3pt}{1em}}%
                     \marginpar{\raisebox{.5em}{\vrule width3pt\ 
                     \vrule width0pt height 0pt depth0.5em 
                     \hbox to 0cm{\hspace{0cm}{%
                     \parbox[t]{4em}{\raggedright\footnotesize{#1}}}\hss}}}}

\def\calm         {{\cal M}}

\def\zet          {{\mathbb Z}}

\def\ee           {{\it e}} 
\def\ii           {{\it i}}

\def\Re           {{\rm Re\hskip0.1em}} 
\def\Im           {{\rm Im\hskip0.1em}}

\def\sqr#1#2{{\vcenter{\vbox{\hrule height.#2pt   
 \hbox{\vrule width.#2pt height#1pt \kern#1pt 
 \vrule width.#2pt}\hrule height.#2pt}}}}

\renewcommand{\P}{\mathbb P}
\newcommand{\R}{\mathbb R}
\newcommand{\Z}{\mathbb Z}
\newcommand{\G}{\mathcal G}
\newcommand{\C}{\mathbb C}

\newcommand{\Ical}{\mathcal I}

\newcommand{\Fcal}{\mathcal F}

\newcommand{\Ocal}{\mathcal O}
\newcommand{\Wcal}{\mathcal W}

\newcommand{\Ncal}{\mathcal N}

\newcommand{\Gcal}{\mathcal G}
\newcommand{\Mcal}{\mathcal M}

\newcommand{\ep}{\epsilon}
\renewcommand{\t}{\tilde}
\newcommand{\h}{\hat}
\newcommand{\corr}[1]{\left<#1\right>}

\newcommand{\beq}{\begin{equation}}
\newcommand{\eq}{\end{equation}}
\newcommand{\req}[1]{(\ref{#1})}

\catcode`\@=12 

\begin{document} 


\title{ABCD of Beta Ensembles and Topological Strings}

\pubnum{
CERN-PH-TH-2012-189 \\
UCB-PTH-12/11             
}
\date{July 2012}

\author{
Daniel Krefl${}^a$ and Johannes Walcher${}^b$ \\[0.2cm]
\it ${}^a$ Center for Theoretical Physics \\
\it University of California, Berkeley, California, USA\\
\it ${}^b$ Departments of Physics, and Mathematics and Statistics\\
\it McGill University, Montr\'eal, Qu\'ebec, Canada
}

\Abstract{
We study $\beta$-ensembles with $B_N$, $C_N$, and $D_N$ eigenvalue measure and 
their relation with refined topological strings. Our results generalize the familiar 
connections between local topological strings and matrix models leading to $A_N$ 
measure, and illustrate that all those classical eigenvalue ensembles, and their 
topological string counterparts, are related one to another via various 
deformations and specializations, quantum shifts and discrete quotients. We review
the solution of the Gaussian models via Macdonald identities, and interpret them
as conifold theories. The interpolation between the various models
is plainly apparent in this case. For general polynomial potential, we calculate 
the partition function in the multi-cut phase in a perturbative fashion, beyond 
tree-level in the large-$N$ limit. 
The relation to refined topological string orientifolds on the corresponding local 
geometry is discussed along the way. 
}

\makepapertitle

\body

\version\versionno

\vskip 1em

\tableofcontents


\section{Introduction}

Eigenvalue ensembles with $A_N$ measure to a power of $\beta$, \footnote{$A_N$ denotes the
finite Coxeter group, and $\beta$ a positive real number. We review the definitions in 
section \ref{MDensDef}.} widely known just as $\beta$-ensembles, and their relation to 
topological gauge and string theories have been studied extensively in recent years.
The special instance $\beta=1$ of generalized interest is the Dijkgraaf-Vafa relation 
\cite{DV02} between matrix models, supersymmetric gauge theory and the topological string. 
In more recent times, the focus has shifted to the more general situation with arbitrary
$\beta$, which relates the eigenvalue ensembles to $\Omega$-deformed gauge theories,
refined topological string theory \cite{DV09,ACDKV11} and the AGT conjecture \cite{AGT09}.
Here, the equivariant parameters $\ep_i$ of the $\Omega$-deformation, the ensemble parameter
$\beta$ and the string coupling $g_s$ are related via \cite{DV09},
\beq\eqlabel{epbeta}
\ep_1=\sqrt{\beta}g_s\,,\,\,\,\,\,\ep_2=-\frac{1}{\sqrt{\beta}} g_s\,.
\eq

One may note that neither the matrix model nor the topological string at present knows a 
microscopic interpretation for the deformation parameter $\beta$. Rather, the mutual
agreement of results calculated with different schemes, the consistency of the space-time
interpretation via BPS state/instanton counting, as well as the relation with the 
$\Omega$-deformed gauge theory, especially in the Nekrasov-Shatashvili limit \cite{NS09}
give confidence that one should view all these models as integral part of a larger 
interconnected web of theories, thereby in fact defining various notions of quantum 
geometry, such as that of \cite{ACDKV11}.

The prototypical example for much of this is the Gaussian model, with quadratic
potential for the eigenvalues and corresponding, respectively, to 
a deformed conifold target space, 
(refined) Chern-Simons theory \cite{AS11}, as well as the $c=1$ non-critical string 
at radius $R=\beta$ \cite{ghva,DV09}. This Gaussian model also serves as building block
for more general backgrounds.

The purpose of the present paper is to take this logic one step further, and to study
the possible role played by eigenvalue ensembles with other finite group measures,
specifically, $B_N$, $C_N$, and $D_N$. These models, which we will refer to as Macdonald
ensembles, are rather natural, and easily defined, but have been less studied
in the recent topological string/gauge theory literature. The ensembles at $\beta=1$ appeared 
briefly in the context of the Dijkgraaf-Vafa relation to four-dimensional $\Ncal=1$ gauge 
theories with $SO/Sp$ gauge groups and adjoint matter, and the realization of these gauge 
theories as string theoretic orientifolds. Most closely related to the spirit of the present 
work are \cite{ACHKR02} and \cite{IKRSV03}. Due to the nature of the original DV conjecture, 
these studies were essentially confined to tree-level. One of the aims of this work is 
to study the $B_N/C_N$ and $D_N$ eigenvalue ensembles with general $\beta$ beyond 
tree-level in greater detail.\footnote{Since the root systems of $B_N$ and $C_N$ differ 
only in the length of the roots, hence the Haar measures are identical up to an overall 
factor (see for instance \cite{NS04}), it will be sufficient for us to consider only the 
$B_N$ and $D_N$ ensembles.} 

It is then natural to expect that the $B_N$, and $D_N$ Macdonald ensembles with general 
$\beta\neq 1$ are related to a refinement of topological string orientifolds, which was
one of the original motivations for the present work.\footnote{Meanwhile, the refinement 
of topological string orientifolds has been studied, with a different perspective and
motivation, by Aganagic and Schaeffer \cite{AS12}.}
In thinking about the various
pictures, it is however important to remember that these eigenvalue ensembles at $\beta=1$ 
are in general {\it not} identical to the usual $SO$ and $Sp$ matrix models. Rather, the 
latter models provide the microscopic realization of the $A_N$ ensemble at $\beta=2$ and 
$\beta=1/2$, respectively. They are dual to $\Ncal=1$ $SO/Sp$ gauge theory with matter 
in the symmetric/antisymmetric representation. In particular, the orientifold in the 
large-$N$ dual topological string side acts differently on the tree-level geometry 
\cite{LLT03,LL03}. 

On the other hand we have the realization, in the Gaussian model,
of the $\beta$-parameter as the radius of the circle for $c=1$ non-critical string.
There, the orbifold of the $c=1$ CFT at the self-dual radius is indeed equivalent to 
the $R=2$ circle theory. This connection suggests the existence of an {\it entire new} 
branch of topological string/matrix model dual pairs that connects up to the standard 
branch at $\beta=2$. Our work suggests that this is where the $B_N$ and $D_N$ Macdonald
ensembles fit in.

Whereas the duality between $\Ncal=2$ $U(N)$ gauge theory softly broken to $\Ncal=1$, $A_N$ 
eigenvalue ensemble with $\beta=1$, and topological string theory on the (spectral curve) 
geometry at large $N$ has been discussed and checked exhaustively in many works, for general
$\beta$ much less is known. For $B_N$ and $D_N$, even at $\beta=1$, no higher genus 
check of the proposed duality between the eigenvalue ensemble and topological string 
orientifolds has been performed. The power of $\beta$ plays a major role beyond tree-level, 
and hence one might hope to be able to learn something about refined topological string 
theory and orientifolds thereof along the way, which are expected to be related to these 
$\beta$-ensembles in the large $N$ limit.

In fact, under which specific conditions the eigenvalue ensembles for $\beta\neq 1$ relate 
to refined topological string theory in the large $N$ limit has not been pointed out so far 
in the literature, even for the ordinary $A_N$ measure. Some examples where such a 
relation holds in a non-trivial manner where reported in \cite{DV09,ACDKV11}. Although
the calculations of \cite{ACDKV11} were restricted to the cubic ensemble at 1-loop level, 
there is little doubt that the observed correspondence extends to all genus, at least for
the cubic. On the other hand, the Chern-Simons matrix models studied in \cite{BMS10} 
appear to indicate that in general such a relation does not hold. Attempts 
to formulate a refined version of the remodeled B-model of \cite{BKMP07} have also failed 
to our knowledge so far. 

Some of the problems with the general applicability of the $A_N$ type $\beta$-ensemble 
can be traced back all the way to tree-level, that is, to the dual spectral curve geometry 
of the ensemble. This is most clearly visible at hand of the remodeled B-model geometries 
of \cite{M06}: In general, the spectral curve of the eigenvalue ensemble differs from the 
usual B-model target space geometry of the dual topological string and has 
singular points. Singularities are a general indication that refinement, \ie, a 
deformation of the correspondence away from $\beta=1$, will fail. Indeed, singularities 
in the B-model geometry could harbor blow-up modes, which spoil an invariant BPS 
state counting. The corresponding mirror statement is the well-known fact that in 
order to have a well-defined BPS state counting of left and right spin (and not 
just the index), the A-model/M-theory geometry should be rigid (\ie, have no complex 
structure deformations). This leads us to a condition on an $A_N$ type $\beta$-ensemble 
to have a well-defined BPS state counting interpretation. Namely, the spectral curve 
has to be non-singular. In particular, this applies as well to ensembles with polynomial 
potentials, \ie, one has to fill all cuts to ensure that one has a well defined BPS 
index. Under this restriction, the duality of \cite{DV09} has a chance to survive 
the $\beta$-deformation in a quite general setting. Similar considerations apply to 
the $B_N$ and $D_N$ cases, up to some technicality which we will explain in more 
detail in section \ref{PertCalcGeneral}. Confirmation for this expectation will be 
found at hand of $A_N$, $B_N$ and $D_N$ $\beta$-ensembles with quartic potential, 
which appear to be as well compatible with a (refined) topological string 
interpretation, as the free energies fulfill the 1-loop holomorphic anomaly equation.

\medskip

The outline is as follows. In section \ref{MDensDef} we will give the definition of Macdonald 
ensembles with special emphasize on $A_N$, $B_N$ and $D_N$. This is followed by a 
detailed discussion of the large $N$ expansion of the Gaussian partition functions and 
implications thereof for refined topological string orientifolds, in section \ref{Gpartitionf}. 
In section \ref{Gcorrelators} a recursion relation satisfied by Gaussian correlators is 
derived (generalizing \cite{AMM03,MS10}), which constitute an essential ingredient for 
the explicit calculation of the 
multi-cut ensemble partition function, which section \ref{PertCalc} is about. In subsection 
\ref{PertCalcGeneral}, we will give a generalization of the framework of 
\cite{AKMV02,KMT02,AMM03,MS10} to $B_N$/$D_N$, and apply it in section 
\ref{BnDnQuarticEnsemble} to the model with quartic potential. The B-model verification 
of the tree-level and 1-loop results of section \ref{BnDnQuarticEnsemble} will be performed 
in section \ref{Bmodel}. We conclude in section \ref{conc}.  In appendix \ref{appA} the 
explicit results for the $g_s$ expansion of the free energy of the $B_N$ and $D_N$ 
$\beta$-ensemble with quartic potential are attached.

\section{Macdonald ensembles}
\label{MDensDef}

Let $\Gcal$ be a finite group of isometries of $\R^N$ generated by reflections in hyperplanes 
through the origin (\ie, a finite reflection or Coxeter group). Let there be $h$ hyperplanes,
each defined by a condition on $\lambda\in\R^N$ of the form
$$
\sum_{i=1}^N a_{\alpha,i}\, \lambda_i=0\,,
$$
where $a_\alpha\in\R^N$, $\alpha=1,\ldots,h$. The group $\Gcal$ naturally acts on 
the algebra of polynomial 
functions on $\R^N$. The $\Gcal$-invariant polynomials form an $\R$-algebra generated by 
$N$ independent polynomials of degrees $d_i$, $i=1,\ldots, N$.
Normalizing the vectors $(a_{\alpha,i})_i$ via $\sum_{i=1}^N a_{\alpha,i}^2=2$, define 
the particular invariant polynomial
$$
P_\Gcal(\lambda)=\prod_{\alpha=1}^h\sum_{i=1}^N a_{\alpha,i}\,\lambda_i\,.
$$
Macdonald conjectured the integral identity \cite{MD82}
\beq\eqlabel{MacDintegral}
Z_\Gcal(\beta):=\frac{1}{(2\pi)^{N/2}}\int_{\R^N}[d\lambda]\,|P_\Gcal(\lambda)|^{2\beta}\, 
e^{-\frac{1}{2} \sum_{i=1}^N\lambda_i^2}=\prod_{i=1}^N\frac{\Gamma(1+d_i\beta)}{\Gamma(1+\beta)}\,,
\eq
where $[d\lambda]:=\prod_{i=1}^Nd\lambda_i$, $\beta\in\C$ with $\Re\beta>0$. A proof of this 
identity has been given by Opdam \cite{O89,O93}. $Z_\Gcal(\beta)$ is also referred to as 
Macdonald integral. 

For $\Gcal=A_{N-1}$ the Macdonald integral specializes to Mehta's integral. For this work, 
in addition the cases $\Gcal=B_N$ and $\Gcal=D_N$ are of particular interest. These give rise
to the integral identities,
\beq\eqlabel{MDintegrals}
\begin{split}
Z_{A_{N-1}}(\beta)&:=\frac{1}{(2\pi)^{N/2}}\int_{\R^N}[d\lambda]\, \Delta(\lambda)^{2\beta} e
^{-\frac 1 2 \sum_i \lambda_i^2}=\prod_{i=1}^{N}\frac{\Gamma(1+i\beta)}{\Gamma(1+\beta)}\,,\\
Z_{B_N}(\beta)&:=\frac{1}{(2\pi)^{N/2}}\int_{\R^N}[d\lambda]\,\Delta(\lambda^2)^{2\beta}\, 
\prod_{i=1}^N\lambda_i^{2\beta} e^{-\frac 1 2 \sum_i \lambda_i^2}=\prod_{i=1}^{N}
\frac{\Gamma(1+2i \beta )}{\Gamma(1+\beta)}\,,\\
Z_{D_N}(\beta)&:=\frac{1}{(2\pi)^{N/2}}\int_{\R^N}[d\lambda]\, \Delta(\lambda^2)^{2\beta} 
e^{-\frac 1 2 \sum_i \lambda_i^2}=\frac{\Gamma(1+N\beta)}{\Gamma(1+\beta)}
\prod_{i=1}^{N-1}\frac{\Gamma(1+2i\beta)}{\Gamma(1+\beta)}\,,
\end{split}
\eq
where $\Delta(\lambda)$ denotes the usual Vandermonde determinant,
$\Delta(\lambda):=\prod_{i<j}(\lambda_i-\lambda_j)$. We are particularly interested in the
large $N$ limit thereof, see section \ref{Gpartitionf}. 

Viewing the above integrals as partition functions of Gaussian eigenvalue ensembles, it is 
natural to define general Macdonald ensembles by replacing the quadratic term
$\sum_i \lambda_i^2$ with a general ``single-trace'' polynomial potential $\sum_i W(\lambda_i)$.
For $\Gcal=A_{N-1}$ the Macdonald ensemble is identical to the usual $\beta$-ensemble. For 
$\Gcal=B_N$ and $D_N$, and with $\beta=1$, these ensembles correspond to the ones considered 
in \cite{ACHKR02} in the context of the 
Dijkgraaf-Vafa relation (with the additional condition $W(x)=W(-x)$).

It is convenient to parameterize the measure $P_{(b,d)}(\lambda):=P_\Gcal(\lambda)$ with 
$\Gcal$ being $A_{N}$, $B_N$ or $D_N$ as 
\beq\eqlabel{PbdDef}
P_{(b,d)}(\lambda)=\Delta_+(\lambda)^{b+d}\Delta_-(\lambda)\prod_{i=1}^N\lambda_i^b\,, 
\eq
where we defined 
$$
\Delta_\pm(\lambda):=\prod_{i<j}^N(\lambda_i\pm\lambda_j)\,.
$$
In particular $\Delta_-(\lambda)=\Delta(\lambda)$, corresponds to the usual Vandermonde, 
and $\Delta_-(\lambda)\Delta_+(\lambda)=\Delta(\lambda^2)$\,. For $(b,d)=(0,0)$  we get 
$P_{A_N}(\lambda)$, $(1,0)$ yields $P_{B_N}(\lambda)$ and $(0,1)$ results in $P_{D_N}(\lambda)$. 
Hence the Macdonald ensembles $Z_{\Gcal}(\beta)$ with $\Gcal=A_N$, $\Gcal=B_N$ or $\Gcal=D_N$ 
can be treated simultaneously via the ensemble \footnote{The $g_s$ dependence needed to 
match to topological strings will be brought in via $W(x)$ and, if necessary, a rescaling 
of the eigenvalues.}
\beq\eqlabel{ZBnDnDef}
Z_{(b,d)}(\beta)\sim \int[d\lambda] |P_{(b,d)}(\lambda)|^{2\beta}\,e^{-
\sum_{i=1}^N W(\lambda_i)}\,,
\eq
with $P_{(b,d)}(\lambda)$ as defined in \req{PbdDef}.

The expectation value for an operator insertion $\h\Ocal$ is defined as usual as  
$$
\langle{\h\Ocal}\rangle_{(b,d)}:= \int[d\lambda] |P_{(b,d)}(\lambda)|^{2\beta}\h
\Ocal\,e^{-\sum_{i=1}^N W(\lambda_i)}\,.
$$
Trivially, we have $\corr{1}_{(b,d)}=Z_{(b,d)}(\beta)$.

It is instructive to compare the formulas for $B_N$ and $D_N$. The only difference 
is the additional factor of $\prod_{i=1}^N\lambda_i^{2\beta}$ for $B_N$, and can be 
interpreted as follows. We know from the $A_N$ $\beta$-ensemble that the insertion of a 
brane at position $x$ in the dual geometry corresponds 
to the insertion of some power of a determinant factor of $\prod_{i=1}^N(x-\lambda_i)$ 
times an overall classical piece of $\psi_{cl}(x)=e^{W(x)}$ \cite{ADKMV03,ACDKV11}. 
Different powers of the insertion correspond to different types of branes \cite{ACDKV11}. 
Thus, the insertion of a brane at $x$ plus a mirror brane at $-x$ corresponds to an 
operator insertion of 
\beq\eqlabel{BranePairDef}
\h\Psi_\beta(x)\h\Psi_\beta(-x)=\psi_{cl}(x)\psi_{cl}(-x)\prod_{i=1}^N(\lambda_i^2-x^2)^\beta\,.
\eq
This implies that the $B_N$ ensemble can be understood as the $D_N$ ensemble with insertion of
an additional  pair of branes at the origin ($x=0$), \ie,
\beq\eqlabel{BnDnRelation}
Z_{(1,0)}(\beta)\sim\langle{\h \Psi_\beta(0)\h \Psi_\beta(0)}\rangle_{(0,1)}= Z_{(0,1)}(\beta)\, 
\Psi^{2\beta}_{(0,1)}(0)\,,
\eq
where we defined the partition function with $h$ coincident 
($\beta$-) branes 
$\Psi^{h\beta}_{(b,d)}(x)$ in the background parameterized by $(b,d)$ as
\beq\label{ZbraneDef}
\Psi^{h\beta}_{(b,d)}(x):=\frac{\bigl\langle{\bigl(\h \Psi_\beta(x)\bigr)^h
\bigr\rangle}_{(b,d)}}{\corr{1}_{(b,d)}}\,.
\eq

\section{Gaussian ensembles: Large \texorpdfstring{$N$}{N} partition functions}
\label{Gpartitionf}

In this section, we review in some detail the large-$N$ expansions of the Gaussian
partition functions \eqref{MDintegrals}, as well as the various ways that these
enter into the topological string.
 
\subsection{Large \texorpdfstring{$N$}{N} expansions}

\paragraph{$A_N$}

In contrast to the $B_N$ and $D_N$ ensembles, the 't Hooft large-$N$ limit of 
$Z_{A_{N-1}}(\beta)$ 
has been studied extensively in the physics literature. The asymptotic expansion as $N\to\infty$
is related to the ``Schwinger'' integral,
\beq\eqlabel{AnSchwinger}
\log Z_{A_{N-1}}(\beta) 
 \sim \int \frac{dt}{t} 
\frac{\ee^{-\mu t}}{(\ee^{\epsilon_1 t}-1)(\ee^{\epsilon_2 t}-1)}=:\log Z_{A}(g_s,\beta)\,,
\eq
where $\epsilon_1$, $\epsilon_2$ are related to $g_s$, $\beta$ as in \eqref{epbeta},
and we have
\beq\eqlabel{muDefN}
\mu:=\sqrt{\beta} g_s  N\,.
\eq
We note the obvious symmetry of the partition function under 
$\epsilon_1\leftrightarrow\epsilon_2$. 
As $g_s\to 0$, we have the well-known asymptotic expansion
\beq
\eqlabel{tZAexpansion}
\int \frac{dt}{t} 
\frac{\ee^{-\mu t}}{(\ee^{\epsilon_1 t}-1)(\ee^{\epsilon_2 t}-1)}\sim
\sum_{n=0}^\infty \Phi_A^{(n)}(\beta)\, \Bigl(\frac{g_s}{\mu}\Bigr)^{n}\,,
\eq
with certain polynomial expressions $\Phi_A^{(n)}(\beta)$. \footnote{Really, 
$\beta^n\Phi_A^{(n)}$ is a polynomial in $\beta$.}

As is well-known, for $n$ even, the $\Phi_A^{(n)}(\beta)$ specialize at $\beta=1$ 
to give the virtual Euler characteristic of the moduli space $\calm_g$ of genus 
$g=\frac{n}{2}+1$ complex curves \cite{HZ86},
\begin{equation}
\Phi_A^{(2g-2)}(1) = \chi(\calm_g) = \frac{B_{2g}}{2g(2g-2)}\,,
\end{equation}
where $B_{n}$ are the Bernoulli numbers. For $n$ odd, the  $\Phi_A^{(n)}(1)$ vanish.

On the other hand, for $n$ odd, the $\Phi_A^{(n)}(\beta)$ specialize at $\beta=2$
to give the virtual Euler characteristic of the moduli space $\calm_{\tilde g}^O$ of
complex curves of genus $\tilde g=n+1$ with a fixed-point free anti-holomorphic involution
(\ie, certain type of real curves) \cite{CZ91,GHJ01,OV02}, up to a rescaling of $g_s$. 
In string theory language, 
the quotients give unoriented Riemann surfaces with genus $g=\tilde g/2$, ($\tilde g$ 
being even), no boundaries, and one crosscap,
\begin{equation}
\eqlabel{tPsiAn1and2}
\Phi_A^{(2g-1)}(2) =-2^{1/2-g} \chi(\calm_{\tilde g}^O) = -2^{1/2-g}\frac{(2^{2g-2}-2^{-1}) 
B_{2g}}{2g(2g-1)}\,.
\end{equation}
For $n$ even, we have that $\Phi_A^{(2g-2)}(2)=2^{-g} \chi(\Mcal_g)$. Hence, the coefficients 
at $\beta=2$ show the typical structure of an orientifold
\beq\eqlabel{PhiA2}
2^{n/2}\Phi_A^{(n)}(2)=\frac{1}{2}\chi(\Mcal_g)-\chi(\Mcal_{\t g}^O)\,.
\eq
For later reference, note that via making use of \req{AnSchwinger} and \req{PhiA2}, one can 
infer as well an integral representation of the generating function for the 
$\chi(\Mcal_{\t g}^O)$, \ie,
\beq\eqlabel{SchwingerChiO}
\mathcal T(g_s):=\log \frac{Z_A(\sqrt{2}g_s,2)}{\sqrt{Z_A(g_s,1)}}=-\frac{1}{2}\int 
\frac{dt}{t} \frac{e^{-\mu t}}{e^{g_s t}-e^{-g_s t}} 
\sim \sum_{n=0}^\infty \chi(\Mcal_{2n}^O) \left(\frac{g_s}
{\mu}\right)^{2n-1}\,.
\eq
A similar, and in fact related, ``Schwinger" integral has appeared before in the
context of the orientifold constant map contribution \cite{KPW09,K10}.

The fact that the $A_N$ $\beta$-ensemble can be used to interpolate between the Euler
characteristic of moduli spaces of complex and real curves was pointed out in \cite{GHJ01}, 
and interpreted as a geometric parameterization. In particular, it was conjectured that 
the $\Phi^{(n)}_A(\beta)$ themselves should describe the Euler characteristic of some
related moduli space.

Although the appearance of the moduli of real curves is suggestive, a simple closed string
theory interpretation is hampered by the fact that the expansion \req{tZAexpansion} 
for $\beta\neq 1$ contains terms of both even and odd powers of $g_s$. This originates 
from the fact that \eqref{AnSchwinger} is not invariant under $(\epsilon_1,\epsilon_2)\to 
(-\epsilon_1,-\epsilon_2)$, except when $\epsilon_1=-\epsilon_2$. As is by now 
well-appreciated, the additional (quantum) shift
\beq\eqlabel{Anshift}
\mu\rightarrow \mu + \frac{\epsilon_1+\epsilon_2}{2}
\eq
restores that symmetry. We have the asymptotic expansion
\begin{equation}\eqlabel{AnShiftGen}
\int \frac{dt}{t} \frac{\ee^{-\mu t}}
{(\ee^{\epsilon_1 t/2}-\ee^{-\epsilon_1 t/2})(\ee^{\epsilon_2 t/2}-\ee^{-\epsilon_2 t/2})}
\sim
\sum_n \Psi^{(n)}_A(\beta) \Bigl(\frac{g_s}{\mu}\Bigr)^n
\end{equation}
with $\Psi^{(n)}_A(\beta)\equiv 0$ for $n$ odd.
This shifted partition function is also identical to the partition function of the $c=1$
string at radius $R\propto\beta$, originally found in \cite{GK90}.
From the above formulas it is clear that $\Phi_A^{(n)}(1) = \Psi_A^{(n)}(1)$. 

Turning to the topological string, it was discovered long time ago in \cite{ghva},
that the integral \eqref{AnSchwinger} at $\beta=1$, \ie, the $c=1$ string at the self-dual
radius, governs the leading behavior of the B-model topological string in the limit in 
which the target space develops a conifold singularity, as the complex structure parameter
$\mu\to 0$. As explained for instance in \cite{HKQ06}, the coefficients 
$\Phi_A^{(n)}(1)=\Psi_A^{(n)}(1)$ 
therefore provide universal boundary condition for solving the topological string via holomorphic
anomaly equation \cite{BCOV93}. As shown in \cite{KW10,KW10b}, see also \cite{HK10},
the one-parameter deformation $\Psi_A^{(n)}(\beta)$ provides the analogous boundary
conditions for solving the refined topological string in the B-model via the same
holomorphic anomaly equation. (Alternatively, one may use the extended holomorphic anomaly
equation of \cite{W07} with boundary conditions provided by the $\Phi_A^{(n)}(\beta)$
to solve for the refined topological string amplitudes after undoing the quantum shift.)
This observation confirms the identification of $\beta$ as the radius $R$ of $c=1$ string
\cite{DV09}.

A seemingly unrelated observation is the fact that the coefficients $\Psi_A^{(n)}(2)$
also have a topological string interpretation, in the context of the real topological
string \cite{KW09}. Namely, writing \cite{KW10}, 
\begin{equation}
\eqlabel{psiA2viaKB1}
2^{n/2}\Psi_A^{(n)}(2) = \frac{1}{2}\bigl(\Psi_A^{(n)}(1) + \Psi_{\rm KB}\bigr)\,,
\end{equation}
the $\Psi_{\rm KB}$ control the leading behavior of the topological string amplitude 
on a genus $g$ Klein-bottle (an unoriented Riemann surface with genus $g$ and 
even number of crosscaps) around a conifold point in moduli space.
This relation begs for a topological interpretation of the $\Psi_{\rm KB}$ similar
to that of the $\Psi_A^{(n)}(1)$ in terms of the moduli space of genus $g$ complex 
curves (with $n=2g-2$). We note that whatever this interpretation is, 
$\Psi_A^{(n)}(1)$ is {\it not} the virtual Euler characteristic of moduli of complex 
curves with fixed-point free anti-holomorphic involution of odd genus $\tilde g=2g-1$
(which are the
covers of these higher genus Klein bottles) studied in \cite{GHJ01}, which vanishes,
but should be closely related to it. It is also interesting to note that the quantum
shift \eqref{Anshift} transforms Klein-bottle contributions into cross-cap contributions, as is
apparent via comparing \req{PhiA2} and \req{psiA2viaKB1}.

\paragraph{$B_N$}

It follows from \req{MDintegrals} that $Z_{B_N}(\beta)$ can be expressed in terms of 
$Z_{A_{N-1}}(\beta)$ as
$$
\log Z_{B_N}(\beta)=\log Z_{A_{N-1}}(2\beta)+N\log\frac{\Gamma(1+2\beta)}{\Gamma(1+\beta)}\,.
$$ 
Since in our 't Hooft limit, we neglect the most singular terms, of positive power of $N$,
we simply write,
\beq\eqlabel{ZBviaZA}
Z_B(g_s,\beta)\sim Z_{A}(g_s,2\beta)\,.
\eq
As a result, the $\beta\leftrightarrow 1/\beta$ symmetry of $Z_A(\beta)$ translates to a 
$\beta\leftrightarrow1/(4\beta)$ symmetry of the $Z_{B}(\beta)$ partition function. Similarly
as in the $A_N$ case, we denote the expansion coefficients of the $g_s$ expansion of the 
corresponding (shifted) free energy as $\Psi_B^{(n)}(\beta)$. Obviously, 
$$
\Psi_B^{(n)}(\beta)=\Psi_A^{(n)}(2\beta)\,.
$$
For later reference, let us explicitly state the ``1-loop" coefficient which the (shifted) 
$Z_B(g_s,\beta)$ implies, \ie, 
\beq\eqlabel{BnPsis}
\begin{split}
\Psi_B^{(0)}(\beta)=\frac{1}{48}\left(\frac{1}{\beta}+4\beta\right)\,.
\end{split}
\eq

\paragraph{$D_N$}

For the $D_N$ Macdonald ensemble, we sort terms such that 
\begin{equation}
\log Z_{D_N}(\beta/2)=
\log Z_{A_{N-1}}(\beta)
-\log\frac{\Gamma(1+N\beta)}{\Gamma(1+N\beta/2)}+N\log\frac{\Gamma(1+\beta)}
{\Gamma(1+\beta/2)}\,.
\end{equation}
Thus, besides the $A_N$ term, we have one additional $N$ dependent term which 
we don't neglect in the large-$N$ limit.  Invoking the integral representation of the digamma 
function $\gamma(x)=\frac{d}{dx}\log(\Gamma(x))$,
$$
\gamma(x)=\int_0^\infty dt\left(\frac{e^{-t}}{t}-\frac{e^{-xt}}{1-e^{-t}}\right)\,,
$$ 
one can infer that the essential part of the new contribution reads
$$
-\int \frac{dt}{t}\frac{e^{-N\beta t}-e^{-N\beta t/2}}{1-e^{-t}}\,.
$$
Redefining $t\rightarrow t \frac{g_s}{\sqrt{\beta}}$ and taking $N\rightarrow\infty$ while 
keeping $\sqrt{\beta}N g_s=:\mu$ fixed yields
$$
-\int\frac{dt}{t}\frac{e^{-\mu t}-e^{-\mu t/2}}{1-e^{-\frac{g_s}{\sqrt{\beta}}t}} \,.
$$
Reverting to our usual notation, we obtain (up to non-universal terms)
\beq\eqlabel{ZDnSchwinger}
\log  Z_{D_N}(\beta/2)\sim \int \frac{dt}{t} \frac{\ee^{-\mu t}}{(\ee^{\epsilon_1 t}-1)
(\ee^{\epsilon_2 t}-1)} + 
\int \frac{dt}{t} \frac{\ee^{-\mu t}}{\ee^{\epsilon_2 t} - \ee^{-\epsilon_2 t}}=:\log 
Z_{D}(g_s,\beta/2)\,.
\eq
We recognize the second term as being essentially the generating function \req{SchwingerChiO}. 
Hence, $Z_D(g_s,\beta)$ is entirely given by a combination of $Z_A$, \ie, 
\beq\eqlabel{ZDwithT}
\log Z_D(g_s,\beta)=\log Z_A(g_s,2\beta)-2\,\mathcal T(\sqrt{2\beta}g_s)\,.
\eq
(where $\mathcal T$ is defined in \eqref{SchwingerChiO}.)
It is also instructive to express the ``1-loop" coefficient $\Phi^{(0)}_D(\beta)$ contained 
in \req{ZDnSchwinger} in terms of $\Phi^{(0)}_A(\beta)$. Using the relations \req{ZDwithT} and 
\req{SchwingerChiO}, we infer
\begin{equation}
\eqlabel{weinfer}
\Phi_D^{(0)}(\beta)=\Phi_A^{(0)}(2\beta)+2\left(\Phi_A^{(0)}(2)-\frac{1}{2}
\Phi_A^{(0)}(1) \right)\,.
\end{equation}

We observe that this result matches the structure of the orbifold branch partition 
function for the $c=1$ CFT on the torus derived in \cite{G87}. We take this as a hint 
that \req{ZDwithT} is related to the partition function of the orbifold branch of the $c=1$
non-critical string. More precisely, such a relation should hold after an appropriate
quantum shift. One way to identify the appropriate $D_N$-analog of \eqref{Anshift} is
to impose a symmetry under $g_s\to-g_s$. The symmetry can be motivated as follows. We 
know that for integer values of $\beta$ the partition function of the $c=1$ string on 
the circle branch can be matched to the partition function of the topological string 
expanded near a $A_{\beta-1}$ type singularity (under appropriate choice of deformation 
parameters) \cite{GV98}. Since the chiral ground ring manifold of the c=1 string on 
the orbifold branch corresponds to a Kleinian singularity of $D$-type \cite{GJM92}, 
we expect that similarly the orbifold branch partition function can be matched to the 
topological string expanded near a $D$-type singularity, implying the symmetry under 
$g_s\to-g_s$.  

Indeed, after
\begin{equation}
\eqlabel{Dnshift}
\mu\rightarrow \mu-\frac{\epsilon_2}{2}\,,
\end{equation}
the free energy \req{ZDnSchwinger} becomes
\beq\eqlabel{TrueDn}
\int\frac{dt}{t}\frac{e^{-\mu t}\cosh\left(\frac{(\ep_1+\ep_2)t}{2}\right)}{2\sinh
\left(\ep_1 t/2\right)\sinh\left(\ep_2 t\right)}\sim\sum_{n}\Psi^{(n)}_D(\beta/2)
\left(\frac{g_s}{\mu}\right)^n\,,
\eq
and clearly possesses an even power only expansion in $g_s$. \footnote{It is interesting 
to note that this generating function looks very similar to the generating function for 
the massless hypermultiplet contribution occuring in SU(2) gauge theory on 
$\Omega$-deformed $A_1$ ALE space \cite{KS11}.} We denote the
expansion coefficients by $\Psi_D^{(n)}(\beta)$. 
For the special value $\beta=1$, we find in addition
\beq
\eqlabel{BDPsiRelationBeta1}
\Psi^{(n)}_D(1)=\Psi^{(n)}_B(1)=\Psi^{(n)}_A(2)\,.
\eq
This identity is precisely the one expected from the matching of circle and orbifold
branches of $c=1$ string at $R=1$ and $R=2$, respectively.
It is important to note however that generally, $Z_{A_N}$ and $Z_{D_N}$ are not related by
a simple shift of $N$.

For later reference, we explicitly state the ``1-loop" coefficient
\beq\eqlabel{DnPsis}
\Psi^{(0)}_D(\beta)=\frac{1}{48}\left(\frac{1}{\beta}-8\beta\right)+\frac{1}{4}\,.
\eq

\subsection{Implications for toric Calabi-Yau backgrounds}

The free energy $\Fcal_{A}(Q;\beta)$ of the refined topological string on the 
{\it resolved} conifold geometry, 
\ie, $\Ocal(-1)\oplus\Ocal(-1)\rightarrow \P^1$, is related to the refined {\it deformed} 
conifold 
free energy given by the integral \req{AnShiftGen} by identifying the K\"ahler
parameter as $Q=e^{-\mu}$ and simply replacing the integral by a sum
\beq\eqlabel{Cquant}
\int dt \rightarrow \sum _{d=1}^\infty\,.
\eq
This replacement originates from the sum over states of D0-brane charge $k\in\zet$
and mass $\sim\mu+2\pi\ii k$, or, in M-theory language, from the extra state degeneracy 
due to momenta around the M-theory circle. It is natural to assume that a similar
``quantization'' as in \eqref{Cquant} can be applied to the other Macdonald integrals
as well. For $B_N$, we infer from \req{ZBviaZA}
\beq\eqlabel{tFBConi}
\Fcal_B(Q;\beta)=\Fcal_A(Q;2\beta)\,,
\eq
\ie, the resolved conifold free energy of $B_N$ type agrees with that of type $A_N$.

The $D_N$ ensemble is more interesting. Applying \req{Cquant} to the (shifted) $D_N$ 
free energy \req{TrueDn}, we obtain
\beq\eqlabel{FDconi}
\Fcal_D(Q;\beta/2)=\sum_{d=1}^\infty Q^d\frac{(q/t)^{d/2}+(q/t)^{-d/2}}
{d(q^{d/2}-q^{-d/2})(t^{-d}-t^{d})}\,,
\eq
with the usual definitions $q:=e^{\sqrt{\beta}g_s}$  and $t:=e^{\frac{g_s}{\sqrt{\beta}}}$, 
as prediction for the refined free energy of $D_N$ type of the resolved conifold geometry.
In particular, we have an even power only expansion in $g_s$ and the relation $\Fcal_D(Q;1)
=\Fcal_A(Q;2)$ holds.

It is instructive to compare this result with the expectations based on a topological
string orientifold interpretation. A convenient reference is the recent proposal
\cite{AS12} for the orientifolded and refined resolved conifold free energy. This
proposal, which is obtained from an $SO(2N)$ refined Chern-Simons/geometric transition
point of view, reads,
\beq\eqlabel{tFDConi}
\Fcal_A(t^{-1/2}Q;\sqrt{2}g_s,2\beta)+\sum_{d=1}^\infty\frac{((q/t)Q)^{d/2}}{t^{d/2}-t^{-d/2}}\,.
\eq
(we have here exchanged $q\leftrightarrow t$ (corresponding to 
$\beta\leftrightarrow 1/\beta$).)
As observed in \cite{AS12}, the specialization of \eqref{tFDConi} to $\beta=1$ equals the free 
energy of an orientifold of the resolved conifold (acting either in fixed-point
free \cite{AAHV02}, or in a real \cite{W072,KPW09} fashion). The second term in 
\eqref{tFDConi} can be understood as originating from the second term in \eqref{SchwingerChiO}
of the (unshifted) $D_N$ free energy, summing only over even D$0$ brane charge (up to a 
shift). It may also be seen as a brane placed at $-1/2\log Q$ in the A-model geometry. 
Since the brane is localized in two space-time dimensions, it is exposed only to a single 
parameter of the $\Omega$-deformation (after a suitable redefinition of parameters). 

The structure of the refined orientifold free energy 
\req{tFDConi} is consistent with the results of \cite{KW10}, where it was found that the 
free energy of the fixed-point free orientifold of $\Ocal(-2)\oplus\Ocal(-2)\rightarrow 
\P^1\times\P^1$ equals the refined free energy at $\beta=2$ (for this orientifold one 
has no open string sector). 

We draw attention to the fact that, at $\beta=1$, the shift of $\log Q$ in the first
term of \eqref{tFDConi} cancels the part of the sum in the second term coming from
even $d$. That summation would be unusual for the $g_s$-odd sector of an orientifold. 
The cancellation is possible because the open string contribution is essentially a closed 
string period. On the other hand, this kind of comparison challenges the extrapolation 
of the proposed orientifold structure \req{tFDConi} to more general toric Calabi-Yau 
geometries, since a cancellation between a shift and an open string contribution is 
not possible in general.
 
Instead, we propose to view $\Fcal_D(Q;\beta)$ of \req{FDconi} as the refined free energy of 
type $D_N$ of the resolved conifold, independent of an orientifold interpretation.\
\footnote{It is conceivable that this result (instead of \req{tFDConi}) can be obtained 
also from the Chern-Simons point of view by appropriately incorporating the quantum shift 
\req{Dnshift} in the large-$N$ limit.}
Although for specific situations (such as the conifold or Dijkgraaf-Vafa type geometries, 
see section \ref{PertCalc}),
when the orientifold contribution is a closed period, the (unshifted) $D_N$ free energy 
can be matched and interpreted at $\beta=1$ as an orientifold, this relation is not 
expected to persist in general.

Specifically, we propose to identify the theory of $D_N$ type with the orbifold 
branch in the $c=1$ moduli space, extending the identification of $A_N$ with the
circle branch of $c=1$ \cite{DV09}. The relation $\beta=R$ is the same on both
branches. For a toric Calabi-Yau manifold, one may then use \req{TrueDn} as boundary
condition on the holomorphic anomaly equation at the conifold point in moduli space
in order to obtain predictions for the $D_N$ theory. We will not pursue this quite 
interesting toric direction further in this work, but rather stick to the Macdonald 
ensemble setting, where the correspondence with orientifolds holds provided we
work with even potentials. This will give further evidence via explicit calculations
that the deformation $\req{ZBnDnDef}$ away from $\beta=1$ is consistent.

\section{Gaussian ensembles: Correlators}
\label{Gcorrelators}

In order to evaluate perturbatively $\beta$-ensembles with multi-cut support we will need 
to evaluate normalized Gaussian correlators defined as
\beq\eqlabel{GcorrDef}
C^{(b,d)}_{k_1,k_2,\dots,k_m}(\beta):=\frac{\corr{\prod_{i=1}^m S_{k_i}}_{(b,d)}}
{\corr{1}_{(b,d)}}=\frac{1}{Z_{(b,d)}(\beta)}\int [d\lambda]|P_{(b,d)}(\lambda)|^{2\beta}
e^{-\frac{1}{2}\sum_{i=1}^N\lambda_i^2} \prod_{i=1}^m S_{k_i}\,,
\eq
with $S_{k}:=\sum_{i=1}^N \lambda_i^k$\, and normalized via the Gaussian partition function  
$Z_{(b,d)}$ given in \req{MacDintegral}. Clearly,
$$
C^{(b,d)}_{0,0,\dots,0}(\beta)=N\times N\times\dots\times N\,.
$$
Correlators with non-vanishing $k_i$ can be solved for recursively invoking the 
Ward-identities resulting from invariance under eigenvalue reparameterizations. 
For example, more recently this approach has been followed in the $A_N$ case with 
general $\beta$ in \cite{MS10}. The generalization of this approach to the $B_N$ and 
$D_N$ case we are interested in is straight-forward. The Ward identities read
\beq\eqlabel{WardId}
\sum_{k=1}^N \int [d\lambda] \partial_{\lambda_k}\left(\lambda_k^n \left(\Delta_+
(\lambda)^{b+d}\Delta_-(\lambda)\prod_{i=1}^N\lambda_i^b\right)^{2\beta} e^{-\frac{1}{2}
\sum_{i=1}^N \lambda_i^2} S_{k_1}\dots S_{k_m}\right)=0\,.
\eq
Acting with the derivative on each factor and expressing the resulting terms through the 
correlators \req{GcorrDef} yields a recursive equation for them. For brevity, we will 
here explicitly state only the new contributions of $\Delta_+(\lambda)^{b+d}$ and 
$\prod_{i=1}^N\lambda_i^b$ which do not occur in the $A_N$ case. The remaining $A_N$ 
contributions can be deduced similarly.

Using, 
$$
\sum_{k=1}^N\lambda^n_{k}\partial_{\lambda_k}\left(\prod_{i=1}^N\lambda_i^b\right)^{2\beta}
= 2\beta bS_{n-1}\left(\prod_{i=1}^N \lambda_i^b\right)^{2\beta}\,,
$$
the new contribution which only occurs for $B_N$ can be inferred to be simply given by
$$
2\beta b\, C^{(b,d)}_{n-1,k_1,\dots,k_m}\,.
$$
The derivation of the contribution of $\Delta_+(\lambda)^{b+d}$ goes as follows. One rewrites
$$
\sum_{k=1}^N\lambda^n_k\partial_{\lambda_k}\left( \Delta_+(\lambda)^{2\beta(b+d)}\right)=
2\beta(b+d)\Delta_+(\lambda)^{2\beta(b+d)}\sum_{k=1}^N\lambda^n_k\partial_{\lambda_k}\log 
\Delta_+(\lambda)\,,
$$
and deduces the identity
$$
2\beta(b+d)\sum_{k=1}^N\lambda^n_k\partial_{\lambda_k}\log \Delta_+(\lambda)=(b+d)\,\beta
\sum_{i\neq j}\frac{\lambda_i^n+\lambda_j^n}{\lambda_i+\lambda_j}=(b+d)\,\beta
\sum_{i\neq j}\frac{\lambda_i^n-(-\lambda_j)^n}{\lambda_i-(-\lambda_j)}\,.
$$
Note that the second step in the above equality is only valid for $n$ odd. Fortunately, 
knowing only how to deal with the $n$ odd case is sufficient for our purposes. Finally, 
making use of the identity
$$
\sum_{i\neq j}\frac{\lambda_i^n-(-\lambda_j)^n}{\lambda_i-(-\lambda_j)}=\sum_{k=0}^{n-1}
\sum_{i\neq j}(-1)^k \lambda_i^{n-k-1}\lambda_j^k=\sum_{k=0}^{n-1}(-1)^kS_{n-k-1}S_k - S_{n-1}\,,
$$
one arrives at the contribution 
$$
\beta(b+d)\sum_{k=0}^{n-1}(-1)^kC^{(b,d)}_{n-k-1,k,k_1,\dots,k_m}-\beta(b+d)
C^{(b,d)}_{n-1,k_1,\dots,k_m}\,.
$$
Combining all terms, taking also the usual $A_N$ contribution into account (see for 
instance \cite{MS10,MMPS11}), the Ward identities \req{WardId} translate to the 
recursive relation
\beq
\begin{split}
C^{(b,d)}_{n+1,k_1,k_2,\dots,k_m}=&\,((1-\beta)n+(b-d)\beta)C^{(b,d)}_{{n-1},k_1,\dots,k_m}\\
&+\beta\sum_{k=0}^{n-1}(1+(b+d)\, (-1)^k)C^{(b,d)}_{{n-k-1},k,k_1,\dots,k_m}+\sum_{j=1}^m  
k_{j} C^{(b,d)}_{{k_1},\dots, {k_j+n-1},\dots, {k_m}}.
\end{split}
\eq
Note that the sole difference between the $B_N$ and $D_N$ case is a switch of sign of one 
of the terms entering the recursive relation. 

As pointed out already above, in case of $B_N$ and $D_N$ the recursive relation is only 
valid for $n$ odd. However, the relation closes if furthermore all $k_i$ are even. Hence, 
it can be used to determine all correlators $C^{(b,d)}_{k_1,k_2,\dots, k_m}$ with $k_i$ 
even. A few examples follow.
\beq
\begin{split}
C^{(0,1)}_{2}&=N(1+2\beta(N-1))\,,\,\,\,\,\,C^{(1,0)}_{2}=N(1+2\beta N)\,,\\
C^{(0,1)}_{2,2}&=C^{(0,1)}_{2}(2+C^{(0,1)}_{2})\,,\,\,\,\,\,C^{(1,0)}_{2,2}=C^{(1,0)}_{2}
(2+C^{(1,0)}_{2})\,,\\
C^{(0,1)}_{4}&=C^{(0,1)}_{2}(3+4\beta(N-1))\,,\,\,\,\,\,C^{(1,0)}_{4}=C^{(1,0)}_{2}
(3+\beta(4N-2))\,.
\end{split}
\eq 

One should note that while in the $A_N$ case with $\beta=1$ a generating function for 
the 1-point correlators $C_{n}^{(0,0)}$ is known \cite{HZ86} and there are also closed 
expressions for $\beta=1/2$ and $\beta=2$ \cite{GJ97}, no such closed formula has been 
found for general $\beta$, nor for the $B_N$ and $D_N$ cases, so far. Nevertheless, we 
can make at least one general observation regarding the structure of the $C_{n}^{(b,d)}$ 
for $n$ even. Namely, the coefficients of the highest powers in $N$ appear to be always 
given in terms of the Catalan-numbers $C_n:=\frac{(2n)!}{(n+1)!n!}$, \ie,
\beq
\begin{split}
C_{2n}^{(0,0)}(\beta)= C_n \beta^n N^{n+1}&+\dots\,,\,\,\,\,\,C_{2n}^{(1,0)}(\beta)= 
C_n \beta^n N^{n+1}+\dots\,,\\
C_{2n}^{(0,1)}(\beta)&= 2^nC_n \beta^n N^{n+1}+\dots\,.
\end{split}
\eq

\section{Multi-cut potentials: Perturbative calculation}
\label{PertCalc}

In this section we shall evaluate the eigenvalue ensembles \req{ZBnDnDef} in 
a perturbative fashion via a saddle-point approximation, making use of the fact that 
for $g_s\rightarrow 0$ the eigenvalues $\lambda_i$ localize to the critical points 
of the potential $W(x)$. A detailed exposition of the perturbative calculation
of the $\beta$-deformed $A_N$ partition function has been given in
\cite{MS10}, as a generalization of the earlier $\beta=1$ works \cite{AKMV02,KMT02,AMM03}. 
We will not repeat that discussion here, but focus on the new features
that appear for $B_N$ and $D_N$ ensembles. We discuss only potentials with the
symmetry $W(x)=W(-x)$. In particular, $W(x)$ is a polynomial of even degree.

\subsection{Saddle-point approximation}
\label{PertCalcGeneral}

Since the degree $d$ of $W(x)$ is even, we have an odd number $c=d-1$ of critical 
points. In particular, 
due to the $\Z_2$ symmetry of the potential there is one critical point at $\Im(x)=\Re(x)=0$. 
We denote the set of critical points as $\mu^{(k)}$ with $k\in\{-\frac{c-1}{2},...,0,...,
\frac{c-1}{2}\}$, hence $\mu^{(-k)}=-\mu^{(k)}$. The saddle-point approximation requires 
us to distribute the $N$ eigenvalues between the $c$ critical points. Let us denote the 
number of eigenvalues located around $\mu^{(k)}$ as $N_k$. In contrast to the $A_N$ case, 
we have to impose some additional constraints onto the eigenvalue distribution. This
will allow the $B_N$ and $D_N$ ensembles to be dual to, both, $\Ncal=2$ $SO$/$Sp$ gauge 
theories with adjoint broken to $\Ncal=1$ by a tree-level potential of the form $W(x)$ 
and to topological 
string orientifolds. From an orientifold point of view, it is more convenient to work in 
the quotient space perspective. In particular, for the eigenvalue ensembles this allows to 
avoid to deal with the interactions between ``mirror" eigenvalues under the $\Z_2$ 
identification of cuts which the duality to orientifolds requires.  Following 
\cite{IKRSV03}, we implement the quotient into the eigenvalue ensemble by localizing 
the eigenvalues around the quotient set of 
critical points, \ie, we take the eigenvalues to be localized around $(c+1)/2$ of the 
critical points such that
$$
N=N_0+...+N_{\frac{c-1}{2}}\,.
$$
The $A_N$ condition for consistency of the $\beta$-deformation of filling all cuts stated 
in the introduction then changes in the $B_N/D_N$ case to filling only the quotient set 
of cuts. 

The partition function can then be evaluated by considering small fluctuations $y^{(k)}_n$, 
with $k\in\{0,\dots,\frac{c-1}{2}\}$, around the critical points, \ie, we set
$$
(\lambda_1,\lambda_2,\dots,\lambda_N)=\left(\mu^{\left(0\right)}+y^{\left(0\right)}_{1},
\dots,\mu^{\left(0\right)}+y_{N_0}^{\left(0\right)},\dots\right)\,.
$$
Under this decomposition we have that
\beq
\begin{split}
\Delta_-(\lambda)&\rightarrow \prod_{k=0}^{\frac{c-1}{2}}\prod_{i<j}^{N_k}\left(y_i^{(k)}-
y_j^{(k)}\right)\prod_{0\leq m<n\leq \frac{c-1}{2}} \prod_{i=1}^{N_m}\prod_{j=1}^{N_n} 
\left(\mu^{(m)}-\mu^{(n)}+y^{(m)}_i-y^{(n)}_j \right)\,, \\
\Delta_+(\lambda)&\rightarrow  \prod_{k=0}^{\frac{c-1}{2}}\prod_{i<j}^{N_k}\left(2\mu^{(k)}
+y_i^{(k)}+y_j^{(k)}\right)\prod_{0\leq m<n\leq \frac{c-1}{2}} \prod_{i=1}^{N_m}
\prod_{j=1}^{N_n} \left(\mu^{(m)}+\mu^{(n)}+y^{(m)}_i+y^{(n)}_j \right)\,.
\end{split}
\eq
Hence, 
\beq\eqlabel{PbdSplit}
|P_{(b,d)}(\lambda)|^{2\beta}\rightarrow \left(\prod_{k=1}^{\frac{c-1}{2}}\Delta_-(y^{(k)})
\Delta_-(y^{(0)})\left(\Delta_+(y^{(0)})\right)^{b+d}\prod_{i=1}^N \left(y_i^{(0)}
\right)^b \right)^{2\beta}\,  \exp \Ical(y)\,,
\eq
with interaction term
\beq\eqlabel{BNDNInterm}
\begin{split}
\Ical(y)=&-2\beta\sum_{0\leq m<n\leq \frac{c-1}{2}}\sum_{l=1}^\infty\sum_{r=0}^l
\frac{(-1)^{r}}{l(\mu^{(m)}-\mu^{(n)})^l}\binom{l}{r}S_r^{(m)}S_{l-r}^{(n)}\\
&-\beta(b+d)\sum_{k=1}^{\frac{c-1}{2}}\sum_{l=1}^\infty\frac{(-1)^l}{2^{l} l 
\left(\mu^{(k)}\right)^l}\sum_{r=0}^l\binom{l}{r}S_r^{(k)}S_{l-r}^{(k)}+\beta(d-b)
\sum_{k=1}^{\frac{c-1}{2}}\sum_{l=1}^\infty\frac{(-1)^l}{l\left(\mu^{(k)}\right)^l}S^{(k)}_l\\
&-2\beta(b+d)\sum_{0\leq m<n\leq \frac{c-1}{2}}\sum_{l=1}^\infty\frac{(-1)^l}
{l(\mu^{(m)}+\mu^{(n)})^l}\sum_{r=0}^l \binom{l}{r}S_r^{(m)}S_{l-r}^{(n)}+{\rm const.}\,.
\end{split}
\eq
The potential decomposes as 
\beq\eqlabel{Wdecompose}
\sum_{i=1}^NW(\lambda_i)\rightarrow\sum_{k=0}^{\frac{c-1}{2}}\sum_{n=1}^\infty 
\frac{(\partial^n W)(\mu^{(k)})}{n!} S_n^{(k)}+{\rm const.}\,.
\eq
Note that the $B_N$, respectively $D_N$ ensemble decomposes in the saddle-point 
approximation into $(c-1)/2$ $A_N$ and a single $B_N$, respectively $D_N$ eigenvalue 
ensembles, which are coupled via the interaction term $\Ical(y)$, as expected. Hence, 
after an appropriate $g_s$ dependent redefinition of $S^{(m)}_i$, and expansion in 
$g_s$ to bring down powers of $S^{(m)}_i$, the partition function of the eigenvalue 
ensemble \req{ZBnDnDef} reduces in the saddle-point approximation to a sum over Gaussian 
correlators which are determinable via the results of section \ref{Gcorrelators}. Due 
to the normalization of the Gaussian correlators, the resulting partition function has 
to be supplemented for each cut by a factor of $Z_\Gcal(\beta)$ (defined in 
\req{MacDintegral}) with $\G$ either $B_N$ or $D_N$ for the fixed cut and else $A_N$.

\subsection{Examples: \texorpdfstring{$B_N$}{B(N)} and \texorpdfstring{$D_N$}{D(N)} quartic} 
\label{BnDnQuarticEnsemble}
Let us now consider an example in more detail. The simplest non-trivial $\Z_2$ 
symmetric example is given by the quartic potential
\beq\eqlabel{quarticpotential}
W(\lambda)=\frac{\beta}{g_s}g\left(\frac{1}{4}\lambda^4-\frac{\delta^2}{2}\lambda^2\right)\,.
\eq
Clearly, $W(-\lambda)=W(\lambda)$ and the set of critical points $\mu^{(k)}$ consists of
$$
\mu^{(-1)}=-\delta\,,\,\,\,\,\,\mu^{(0)}=0\,,\,\,\,\,\,\mu^{(1)}=\delta\,,
$$
hence possesses a three-cut structure. The partition functions $Z_{(b,d)}$ for $A_N$, 
$B_N$ and $D_N$ measure can be obtained in a perturbative fashion as outlined in the 
previous section. While one has to fill in the $A_N$ case all three cuts as discussed 
in the introduction, for $B_N$ and $D_N$ one has to fill only two of the cuts in order 
to incorporate the $\Z_2$ quotient as mentioned in section \ref{PertCalcGeneral}.

For the $A_N$ case, let us just quote the relevant observation without giving any 
further details. Namely, we observe that if we fill all three-cuts, the disk sector 
($g_s^{-1}$) of the corresponding free energy is a combination of closed periods, \ie, 
$$
\t\Fcal_A^{(1/2)}=\frac{1}{2}\left(1-\frac{1}{\beta}\right)\left(\frac{\partial
\Fcal_A^{(0)}}{\partial \t S_{-1}}+\frac{\partial\Fcal_A^{(0)}}{\partial \t S_0}+
\frac{\partial\Fcal_A^{(0)}}{\partial \t S_{1}}\right)\,,
$$
where as usual $\t S_i:=N_i g_s$. This indicates that one has to perform the additional 
quantum shifts
\beq\eqlabel{SiQuant}
S_i:=\left(N_i-\frac{1}{2}\left(1-\frac{1}{\beta}\right)\right) g_s\,,
\eq
in the large $N$ limit. Indeed, after the shifts, one obtains an expansion into 
only even powers of $g_s$ of the free energy $\Fcal_A$, as is necessary for a 
well-defined BPS index interpretation of the corresponding partition function. 

Let us now consider the $B_N$ and $D_N$ cases. For the quartic potential 
\req{quarticpotential}, the measure specializes to
$$
[d y^{(0)}][d y^{(1)}]\left(\Delta_-(y^{(1)})\Delta_-(y^{(0)})\left(\Delta_+(y^{(0)})
\right)^{b+d}\prod_{i=1}^N \left(y_i^{(0)}\right)^b\right)^{2\beta}\,,
$$
the interaction term \req{BNDNInterm} reads
\beq
\begin{split}
\Ical(y)=&-2\beta\sum_{l=1}^\infty\sum_{r=0}^l\frac{(-1)^{l}}{l \delta^l}\binom{l}{r}
\left(1+(b+d) (-1)^r\right)S_r^{(0)}S_{l-r}^{(1)}\\
&-\beta(b+d)\sum_{l=1}^\infty\frac{(-1)^l}{2^{l} l\delta^l}\sum_{r=0}^l\binom{l}{r}
S_r^{(1)}S_{l-r}^{(1)}+\beta(d-b)\sum_{l=1}^\infty\frac{(-1)^l}{l\delta^l}S^{(1)}_l
+{\rm const.}\,,
\end{split}
\eq
and the potential contribution \req{Wdecompose} is given by
$$
\Wcal(y)=-\sum_{n=1}^\infty \frac{1}{n!}\left((\partial^n W)(0) S_n^{(0)}+
(\partial^n W)(\delta) S_n^{(1)}\right)\,.
$$
After performing the rescalings
$$
S_n^{(0)}\rightarrow \left(-\frac{g_s}{\beta g\delta^2}\right)^{n/2}S_n^{(0)}
\,,\,\,\,\,\,S_n^{(1)}\rightarrow \left(\frac{g_s}{2\beta g\delta^2}\right)^{n/2} S_n^{(1)}\,,
$$
the partition functions can be expanded in $g_s$ and reduce to a sum over Gaussian 
correlators, which one can efficiently calculate following section \ref{Gcorrelators}. 
For the reader's convenience, the explicit free energies to some lower order in $g_s$ 
are given in appendix \ref{appA}. Defining
\beq\eqlabel{SiNi}
S_0=2N_0g_s\,,\,\,\,\,\, S_1=N_1g_s,
\eq
we obtain
\beq\eqlabel{tFABD0}
\t\Fcal_{B}^{(0)}(S_0,S_1)=\t\Fcal_D^{(0)}(S_0,S_1)=\frac{1}{2}\t \Fcal_A^{(0)}
(S_0,S_1,S_{-1}=S_{1})\,,
\eq
which is the expected tree-level result from a topological string orientifold 
perspective. The first order open string corrections ($g_s^{-1}$) read
\beq\eqlabel{FBDgsm1}
\begin{split}
\t \Fcal_B^{(1/2)}&=\frac{1}{2}\left(2-\frac{1}{\beta}\right)\frac{\partial
\Fcal_B^{(0)}}{\partial \t S_0}+\frac{1}{2}\left(1-\frac{1}{\beta}\right)
\frac{\partial\Fcal_B^{(1)}}{\partial \t S_1}\,,\\
\t \Fcal_D^{(1/2)}&=-\frac{1}{2\beta}\frac{\partial\Fcal_D^{(0)}}{\partial \t 
S_0}+\frac{1}{2}\left(1-\frac{1}{\beta}\right)\frac{\partial\Fcal_D^{(1)}}{\partial \t S_1}\,,
\end{split}
\eq
and are combinations of closed string periods. Note that for $\beta=1$ we have that 
$\t \Fcal_B^{(1/2)}=-\t \Fcal_D^{(1/2)}
=-\frac{1}{4}\partial_{S_0}\Fcal_A(S_0,S_1,S_{-1}=S_{1})$, confirming the earlier results 
of \cite{ACHKR02,INO02,JO02,IKRSV03}. In the dual topological string orientifold, the sign 
difference translates into the two possible choices of charge of the orientifold 
fixed-plane. Similar as for the $A_N$ case, the order $g_s^{-1}$ given in \req{FBDgsm1} 
suggests to perform the additional quantum shifts as in \req{SiQuant} such that
\beq
\Fcal_B^{(1/2)}=-\frac{1}{2}\frac{\partial\t\Fcal_A^{(0)}(S_0,S_1,S_{-1}=S_1)}{\partial 
\t S_0}\,,\,\,\,\,\,\t \Fcal_D^{(1/2)}=\frac{1}{2}\frac{\partial\t\Fcal_A^{(0)}
(S_0,S_1,S_{-1}=S_1)}{\partial \t S_0}\,,
\eq	
and the relation $ \Fcal_B^{(1/2)}=- \Fcal_D^{(1/2)}$ continues to hold under the 
$\beta$-deformation. In particular, the open string contribution at order $g_s^{-1}$ 
is independent of $\beta$. However, for higher orders in $g_s$, one has that generally
$$
\Fcal_B^{(g>1/2)}(\beta)\neq \Fcal_D^{(g>1/2)}(\beta),
$$
and equality (up to overall sign) only for $\beta=1$. Since the open string 
contribution is trivial (\ie, it is a closed period), it is more convenient to 
shift away the complete open string contribution such that  $\Fcal_B^{(g/2)}= 
\Fcal_D^{(g/2)}=0$, for $g$ odd. Hence, in this specific gauge the free energies 
possess an expansion into even powers of $g_s$ only. If necessary, the open string 
contribution can be easily reinstated via performing a reverse shift. The main 
advantage of this shift is that it allows us to utilize the usual holomorphic 
anomaly of \cite{BCOV93} instead of the extended holomorphic of \cite{W07} to 
reproduce the $B_N$ and $D_N$ partition functions in the B-model. On a technical 
level, the former is easier to deal with. However, we like to stress that the 
latter is more general, since it is expected to capture the partition function 
independent of any shift of parameters, similar as observed at hand of gauge 
theory in \cite{KW10,KW10b}.

Finally, let us comment on the Nekrasov-Shatashvili limit of the free energies 
(in the gauge with an even power $g_s$ expansion). In our parameterization the 
limit of \cite{NS09} corresponds to
\beq\eqlabel{NSlimit}
\Wcal^{(g)}_\Gcal:=\lim_{\beta\rightarrow 0}\beta^{g} \Fcal^{(g)}_\Gcal(\beta)\,.
\eq
From our explicit computations we observe that 
\beq\eqlabel{NSresult}
\frac{1}{2} \Wcal^{(g)}_A=\Wcal^{(g)}_B=\Wcal^{(g)}_D\,.
\eq
A similar non-uniqueness property of the limit has been already observed at hand of gauge 
theory on ALE space in \cite{KS11}, and is in fact as expected. This is because, since 
$g_s\rightarrow 0$ (in order to keep $\hbar:=\frac{g_s}{\sqrt{\beta}}$ fixed, \cf, 
\req{epbeta}), the Nekrasov-Shatashvili limit, and hence $\Wcal^{(g)}$, is intrinsically 
of tree-level nature. More specifically, the limit corresponds to a (semi-classical limit 
of a) quantization of the spectral curve of the respective eigenvalue ensemble, following 
\cite{ACDKV11}. Since the spectral curves of the $A_N$, $B_N$ and $D_N$ ensembles (under 
a proper $\Z_2$ identification of the $A_N$ spectral curve) are identical (\cf, \req{tFABD0}), 
so should be the quantization thereof. The relation \req{NSresult} shows that this is indeed 
the case.

\section{B-model verification of \texorpdfstring{$B_N$}{B(N)} and 
\texorpdfstring{$D_N$}{D(N)} quartic}
\label{Bmodel}

\subsection{Tree-level geometry}

The dual tree-level geometry of the quartic eigenvalue ensemble with $B_N$ and $D_N$ measure 
with $\beta=1$ has been discussed already in the literature to some extent (\cf, 
\cite{FO02,ACHKR02,IKRSV03}). Since the power of $\beta$ is only relevant at one-loop 
and beyond, we essentially can borrow the known tree-level results. 

\begin{figure}[t]
\begin{center}
\psfrag{N}[cc][][0.7]{$0$}
\psfrag{MD}[cc][][0.7]{$-\delta$}
\psfrag{PD}[cc][][0.7]{$+\delta$}
\psfrag{L}[cc][][0.7]{$\Lambda$}
\psfrag{K}[cc][][0.7]{$-\Lambda$}
\psfrag{S0}[cc][][0.9]{$S_0$}
\psfrag{S1}[cc][][0.9]{$S_1$}
\psfrag{S1p}[cc][][0.9]{$S_1'$}
\psfrag{P0}[cc][][0.9]{$\Pi_0$}
\psfrag{P0p}[cc][][0.9]{$\Pi_0'$}
\psfrag{P1}[cc][][0.9]{$\Pi_1$}
\psfrag{P1p}[cc][][0.9]{$\Pi_1'$}
\includegraphics[scale=0.28]{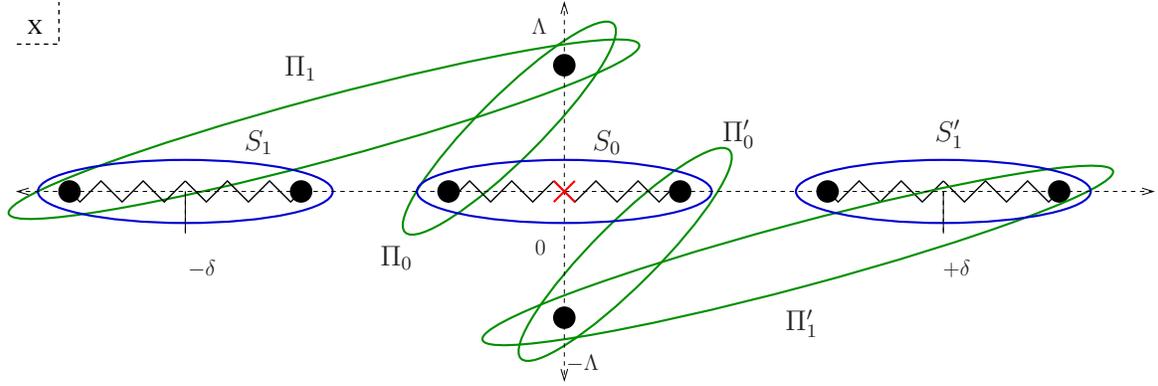}
\caption{The symmetric $x$-plane of the quartic with cuts, period contours and reflection 
symmetry indicated (covering space perspective). }
\label{xplaneQuartic}
\end{center}
\end{figure}

The periods of the dual geometry of the eigenvalue ensemble with potential $W(x)$ of the 
form \req{quarticpotential} are given in terms of the periods of the hyperelliptic curve
\beq\eqlabel{curveEq}
y=M(x)\sqrt{\sigma(x)}\,,
\eq
with
\beq\eqlabel{MsigVars}
M(x)=g\,,\,\,\,\,\,\sigma(x)=\frac{1}{g^2}\left(W'(x)^2+f(x)\right),
\eq
and where $f(x)$ is a degree two polynomial. In particular, the moment function $M(x)$ is 
a constant because we fill all cuts. For the quartic, the curve \req{curveEq} is of genus 
two. The effective one-form of the dual geometry reads
\beq\eqlabel{DV1form}
\omega=y\, dx\,,
\eq
which we express in terms of the six branch points $x_i$ of the curve as 
$$
\omega=g\sqrt{\prod_{i=1}^6(x-x_i)}\,dx\,.
$$
The cuts are chosen to be $[x_1,x_2]$ , $[x_3,x_4]$ and $[x_5,x_6]$ on the real axis. 
Imposing the $\Z_2$ symmetry under $x\rightarrow -x$ of the $x$-plane (this requires 
that $f(x)$ is even, \ie, $f(x)=b_2 x^2+b_0$ with $b_i$ parameterizing the complex 
structure) leads to the identification $x_1\leftrightarrow -x_6$, $x_2\leftrightarrow 
-x_5$ and $x_3\leftrightarrow -x_4$, yielding the one-form 
$$
\omega=g\sqrt{(x^2-x_1^2)(x^2-x_2^2)(x^2-x_3^2)}\,dx\,.
$$
The $\Z_2$ symmetric x-plane of the geometry is illustrated in figure \ref{xplaneQuartic}. 

One should note that if one adjusts such that $b_2=0$, the curve \req{curveEq} takes the 
form of the Seiberg-Witten curve of four dimensional $\Ncal=2$ $SU(3)$ gauge theory, with 
choice of Coulomb parameters $a_1=-a_2$ and $a_3=0$ (the zeroes of $W'(x)$ correspond to 
the $a_i$), which can also be matched to the curve of $Sp(2)$. It would be interesting to 
extract the gauge theory gauge coupling and 1-loop gravitational correction following 
\cite{DGKV02,KMT02} and see if one can match to a $\Omega$-deformed gauge theory, as 
is the case for the cubic and $\Omega$-deformed $SU(2)$ \cite{ACDKV11}. However, since 
$a_3=0$, we are not on the Coloumb branch, and things are expected to be a bit more 
tricky. Therefore we will not follow this rather interesting direction further in this 
work. 

Comparison with the one-form \req{DV1form} expressed via \req{MsigVars} gives the relation 
\beq\eqlabel{Deltaxi}
\delta^2=\frac{1}{2}\left(x_1^2+x_2^2+x_3^2\right)\,.
\eq
In order to explicitly calculate the period integrals, it is useful to change variables to
\beq\eqlabel{zCoords}
z_0=x_3\,,\,\,\,\,\,z_1=\frac{1}{2}(x_2-x_1)\,,\,\,\,\,\,\Ical=\frac{1}{2}\left(x_1+x_2\right)\,.
\eq
Hence, we set
$$
x_3=z_0\,,\,\,\,\,\, x_1=\Ical-z_1\,,\,\,\,\,\, x_2=\Ical+z_1\,.
$$
Using \req{Deltaxi}, we can infer for $\Ical$ the relation
$$
\Ical=\pm\sqrt{\delta^2-\frac{1}{2} z_0^2-z_1^2}\,.
$$
The discriminant $\Delta$ of the algebraic curve \req{curveEq} reads in the $z_i$ 
coordinates \req{zCoords}
\beq
\Delta=4 z_0^2 z_1^4\, \Delta_{1}^2\, \Delta_{2}^2\, \Delta_3^4\,,
\eq
with components
\beq\eqlabel{dissloci}
\begin{split}
\Delta_1&=z_0^2+2z_1^2-2\delta^2\,,\,\,\,\,\,\Delta_2=z_0^2+4z_1^2-2\delta^2\,,\\
\Delta_{3}&=9z_0^4+4z_0^2(2z_1^2-3\delta^2)+4(-2z_1^2+\delta^2)^2\,.
\end{split}
\eq
In particular, we have that $\Delta_1=-2\Ical^2$.

The A-periods of the curve \req{curveEq} with one-form \req{DV1form} are taken to be
\beq
S_0=\frac{1}{2\pi \ii}\int_{-x_3}^{x_3} \omega\,,\,\,\,\,\,S_1=\frac{1}{2\pi \ii}
\int_{x_2}^{x_1} \omega\,.
\eq
It is not hard to explicitly evaluate the integrals in the coordinates \req{zCoords} 
for small $z_i$. The first few terms read
\beq\eqlabel{Sizi}
\begin{split}
S_0(z_i)&=-\frac{g \delta^2}{4}z_0^2+\frac{g}{2} z_0^2 z_1^2+\frac{3g}{16}z_0^4+
\frac{g}{8\delta^2}z_0^4 z_1^2+\frac{g}{4\delta^4}z_0^4 z_1^4+\frac{g}{8\delta^4}
z_0^6 z_1^2+\dots\,,\\
S_1(z_i)&=\frac{g \delta^2}{2}z_1^2-\frac{g}{2} z_0^2 z_1^2-\frac{g}{2}z_1^4-
\frac{g}{16\delta^2}z_0^4 z_1^2-\frac{g}{8\delta^4}z_0^4 z_1^4-\frac{g}{16\delta^4}
z_0^6 z_1^2+\dots\,,
\end{split}
\eq
Note that the two periods are related via the identity
\beq\eqlabel{APeriodRelation}
S_1+\frac{1}{2}S_0=\frac{g}{32}(z_0^2-4z_1^2)(3z_0^2+4z_1^2-4\delta^2)\,.
\eq
Inversion of \req{Sizi} yields the so-called mirror maps $z_i(S_i)$.

Similarly, it is not hard to explicitly evaluate the B-periods 
$$
\Pi_0=\frac{1}{2}\int^\Lambda_{x_3}\omega\,,\,\,\,\,\,\Pi_1=\int^\Lambda_{x_1} \omega\,,
$$
where $\Lambda\rightarrow\infty$ is a cutoff. Note the additional factor of $1/2$ we 
introduced for $\Pi_0$. Its origin can be most easily seen at hand of figure 
\ref{xplaneQuartic}. The period $\Pi_0$ (without the $1/2$) in the quotient 
space corresponds in the covering space actually to $2\Pi_0$ and not just $\Pi_0$.

 We obtain in terms of the flat coordinates $S_i$ for the B-periods
\beq
\begin{split}
\Pi_0(S_i)=&\frac{g}{8}(\Lambda^2-2\delta^2)\Lambda^2-\frac{1}{2}S_0\log g -2(S_1+
\frac{1}{2}S_0)\log\Lambda\\
&+2(S_1-\frac{1}{2}S_0)\log\delta+P_0(S_0,S_1)\,,\\
\Pi_1(S_i)=&\frac{g}{4}(\delta^2-\Lambda^2)^2-4(S_1+\frac{1}{2} S_0)\log\Lambda+
2(S_1+S_0)\log\delta+S_1\log\left(\frac{2}{g\delta^2}\right)\\
&+P_1(S_0,S_1)\,,
\end{split}
\eq
with $P_i(S_0,S_1)=S_i\log S_i+\sum_{n,m\geq 0}\frac{c_i(n,m)}{(g\delta^2)^{n+m-1}} 
S_0^n S_1^m$ and $c_i(n,m)$ constants . 

Invoking the usual special geometry relation
$$
\partial_{S_i}\Fcal^{(0)}=\Pi_i\left(S_0,S_1)\right)\,,
$$
the prepotential $\Fcal^{(0)}$ can be determined, and indeed matches the results of 
section \ref{BnDnQuarticEnsemble}. Note that the $P_i(S_0,S_1)$ can be expressed in 
terms of the flat-coordinate Yukawa couplings $C_{S_iS_jS_k}$ as $P_i(S_0,S_1)=\int 
dS_i dS_i C_{S_i S_i S_i}$. Closed expressions for the Yukawa couplings in $z_i$ 
coordinates, \ie,  $C_{z_i z_j z_k}:=D_{z_i} D_{z_j} D_{z_k} \Fcal^{(0)}(z_i)$ with 
$D_{z_i}$ denoting the covariant derivative (\cf, \cite{BCOV93}), can be found to be
\beq\eqlabel{Yuks}
\begin{split}
C_{z_0 z_0 z_0}&=-\frac{z_0(9z_0^6+6z_0^4(-5+z_1^2)+8(1-2z_1^2)^2(-1+z_1^2)+4z_0^2
(7-8z_1^2+4z_1^4))}{32 \Ical^2}\,,\\
C_{z_1 z_1 z_1}&=-\frac{z_1(3z_0^6+8(-1+2z_1^2)^3+2z_0^4(-7+10z_1^2)+4z_0^2(5-16z_1^2
+12z_1^4)}{4\Ical^ 2}\,,\\
C_{z_0 z_0 z_1}&=\frac{z_1z_0^2(-2+3z_0^2)(-2+z_0^2+4z_1^2)}{8\Ical^2}\,,\\
C_{z_0 z_1 z_1}&=\frac{z_0z_1^2(z_0^4-4(1-2z_1^2)^2)}{4\Ical^2}\,.
\end{split}
\eq
The remaining couplings follow by symmetry. We have also set for simplicity $g=\delta=1$.

\subsection{One-loop}
Having the tree-level data at hand, it is straight-forward to evaluate the solution 
to the 1-loop holomorphic anomaly equation of \cite{BCOV93a} 
\beq\eqlabel{F1Amb}
\Fcal^{(1)}(z;\beta)=\frac{1}{2}\log\det G+a^{(1)}(z;\beta)\,,
\eq
with $G_{ij}:=\partial_{S_i}z_j$ and $a^{(1)}(z;\beta)$ denoting the 1-loop holomorphic 
ambiguity. The ambiguity can be parameterized in terms of the discriminant loci 
$\Delta_i$  given in \req{dissloci} as
$$
a^{(1)}(z;\beta)=\nu_0\log z_0+\nu_1\log z_1+\kappa_1\log\Delta_1+\kappa_2\log\Delta_2
+\kappa_3\log\Delta_3\,.
$$
From the eigenvalue ensemble results of section \ref{BnDnQuarticEnsemble} we deduce 
that under fixing parameters $\nu_i$ and $\kappa_i$ in the $B_N$ case to 
\beq
\begin{split}
\nu_0&=\frac{1}{24}\left(\frac{1}{\beta}+4\beta\right)-\frac{1}{2}\,,\,\,\,\,\,\nu_1
=\frac{1}{12}\left(\frac{1}{\beta}+\beta\right)-\frac{1}{2}\,,\\
\kappa_1&=\nu_0\,,\,\,\,\,\,\kappa_2=\frac{1}{2}\nu_1+\frac{1}{4}\,,\,\,\,\,\,\kappa_3
=\nu_1\,,
\end{split}
\eq
and for $D_N$ to 
\beq
\begin{split}
\nu_0&=\frac{1}{24}\left(\frac{1}{\beta}-8\beta\right)\,,\,\,\,\,\,\nu_1=\frac{1}{12}
\left(\frac{1}{\beta}+\beta\right)-\frac{1}{2}\,,\\
\kappa_1&=\nu_0\,,\,\,\,\,\,\kappa_2=\frac{1}{2}\nu_1+\frac{1}{4}\,,\,\,\,\,\,\kappa_3=\nu_1\,,
\end{split}
\eq
the previous results can be reproduced. We observe that the ambiguities for both cases 
differ in general only in $\nu_0$. Further, note that only in the special case of
$\beta=1$ and in the Nekrasov-Shatashvili limit ($\beta\rightarrow 0$) the $\nu_0$ of 
both cases are equal and the relation
$$
\Fcal^{(1)}_{B}=\Fcal^{(1)}_{D}\,,
$$
holds, as expected from the relations \req{BDPsiRelationBeta1} and \req{NSresult}. For 
general $\beta$ this equality will not hold anymore. Also note that in the 
Nekrasov-Shatashvili limit \req{NSlimit} the 1-loop amplitude \req{F1Amb} becomes 
purely holomorphic, \ie,
$$
\Wcal^{(1)}(z)=\frac{1}{24}\log z_0+\frac{1}{12}\log z_1+\frac{1}{24}\log\left(\Delta_1 
\Delta_2\right)+\frac{1}{12}\log \Delta_3\,.
$$
The reason being that in this limit the 1-loop anomaly equation reduces to 
$$
\bar\partial_{\bar i}\partial_j \Wcal^{(1)}(z,\bar z)=0\,.
$$
To conclude this section, it is interesting to compare the $\nu_0$ coefficients we 
found to the corresponding $\Psi^{(0)}_\Gcal$ coefficients of section \ref{Gpartitionf}, 
given in \req{BnPsis} and \req{DnPsis}. Up to an addition of $1/2$ they match. We 
attribute the additional $1/2$ to an artifact of our expansion at the 1-loop level 
and/or to the chosen parameterization. Mainly because the same mismatch by $1/2$ 
occurs for $\nu_1$, which should be equal to $\Psi^{(0)}_A$.

\section{Conclusion}
\label{conc}

In this work we initiated the study of $\beta$-ensembles with $B_N$ and $D_N$ measure 
beyond tree-level. For that purpose, we generalized the calculation of the 
$\beta$-deformed $A_N$ partition function of \cite{MS10}, which makes use of a 
saddle-point approximation and Ward identities, to the $B_N$ and $D_N$ cases. At 
hand of the quartic, we found that the resulting free energies possess an expansion 
into even powers of $g_s$ only, under a specific choice of 't Hooft parameters. This 
is as expected, since the $g_s^{-1}$ sector is a closed period and should be removable 
via an appropriate shift of parameters following \cite{KW10b}. The absence of an odd 
sector in $g_s$ allowed us to invoke the usual holomorphic anomaly equation to reproduce 
the 1-loop sector ($g_s^0$) of the quartic, albeit with new boundary conditions 
(holomorphic ambiguity) which have not appeared (to our knowledge) before. The 
boundary conditions are related to a large $N$ expansion of the Macdonald integral. 
We expect that the higher genus coefficients of the Macdonald integral expansions 
will provide boundary conditions for the higher loop amplitudes expanded near some 
of the other points in moduli space. However, so far we have not pushed the holomorphic 
anomaly calculation for the quartic beyond genus one, and it would be interesting to do 
so. Our results indicate that the $\beta$-deformation of the $B_N$ and $D_N$ ensembles, 
which for $\beta=1$ correspond to topological string orientifolds, is consistent. 

The Gaussian integrals also allowed us to extrapolate the $B_N$ and $D_N$ ensembles to toric 
settings. We found that the $D_N$ case (under an appropriate shift) should 
correspond to the topological string with boundary conditions at the conifold point 
provided by the orbifold branch of the $c=1$ string. The $B_N$ case corresponds to 
the usual circle branch, similar to the standard refined topological string
related to $A_N$ ensemble. In the toric setting, our results 
indicate that the equivalence between $B_N$/$D_N$ and topological string orientifolds 
is not general. Rather the $c=1$ moduli space appears to yield an independent deformation 
space of topological string theories, with only accidental correspondence to topological 
string orientifolds for specific simple geometries. 

It seems likely that one may also explore ensembles with $E_6$, $E_7$ and $E_8$ type 
measure in a similar fashion. Presumably, these ensembles are related to the three 
discrete points in $c=1$ moduli space (and should not possess a consistent 
$\beta$-deformation).

\acknowledgments

We like to thank S-Y. D. Shih for related discussions. We thank the KITP for
hospitality during the INTEGRAL11 program where this work was started. D.K.\ thanks
McGill University, and J.W.\ thanks CERN-PH TH Division for hospitality
during completion of this work.
The work of D.K. has been supported by a Simons fellowship, and by 
the Berkeley Center for Theoretical Physics. The work of J.W.\ is supported in part by
the Canada Research Chair program and an NSERC discovery grant.
This work was also supported in part by DARPA under Grant No.
HR0011-09-1-0015, and the NSF under Grant No. PHY05-51164.

\newpage
\appendix

\section{Free energies}
\label{appA}

The explicit results for the perturbative expansion of the free energies of the $B_N$ and $D_N$ eigenvalue ensemble with quartic potential discussed in section \ref{BnDnQuarticEnsemble} are given below. Note that the expressions given do not take the normalization of \req{GcorrDef} into account. The additional contribution from the normalization to the shifted free energy is given in terms of an asymptotic expansion of the respective Macdonald integral given in \req{MDintegrals}.

\tiny
\begin{displaymath}
\begin{split}
\t\Fcal_B(N_i,\beta)=&-\frac{g_s}{4\delta^4\beta}\left(8 N_0^3 \beta ^2-2 N_0^2 \beta  (2 \beta  (8 N_1+1)-5)+N_0 \left(8 \beta ^2 (N_1-3) N_1+\beta  (8 N_1-2)+3\right)+(2 \beta -1) N_1 (2 \beta  (N_1-2)+3)\right)\\
&+\frac{g_s^2}{4g^2\delta^8\beta^2}\left(36 N_0^4 \beta ^3-4 N_0^3 \beta ^2 (2 \beta  (28 N_1+5)-19)+N_0^2 \beta  \left(12 \beta ^2 \left(12 N_1^2-16 N_1+1\right)+\beta  (24 N_1-50)+53\right)\right.\\
&\left.+N_0 \left(-8 \beta ^3 N_1
   \left(2 N_1^2-15 N_1+19\right)+\beta ^2 \left(-48 N_1^2+200 N_1+6\right)-\beta  (88 N_1+15)+12\right)-(2 \beta -1) N_1 \left(4 \beta ^2 \left(N_1^2-5 N_1+5\right)\right.\right.\\
   &\left.\left.+\beta (15 N_1-29)+12\right)\right)\\
   &-\frac{g_s^3}{12g^3\delta^{12}\beta^3}\left(864 N_0^5 \beta ^4-8 N_0^4 \beta ^3 (4 \beta  (233 N_1+49)-331)+8 N_0^3 \beta ^2 \left(8 \beta ^2 \left(131 N_1^2-102 N_1+16\right)-4 \beta  (29 N_1+103)+383\right)\right.\\
   &\left.-4 N_0^2
   \beta  \left(4 \beta ^2 \left(165 N_1^2-888 N_1-80\right)+4 \beta ^3 \left(152 N_1^3-558 N_1^2+640 N_1+15\right)+2 \beta  (965 N_1+286)-393\right)+N_0 \left(8 \beta ^2
   \left(419 N_1^2\right.\right.\right.\\
   &\left.\left.\left.-798 N_1+48\right)+32 \beta ^4 N_1 \left(5 N_1^3-66 N_1^2+206 N_1-165\right)+8 \beta ^3 \left(112 N_1^3-1050 N_1^2+1208 N_1-15\right)+12 \beta  (88
   N_1-43)+297\right)\right.\\
   &\left.+(2 \beta -1) B \left(4 \beta ^2 \left(66 N_1^2-314 N_1+307\right)+8 \beta ^3 \left(5 N_1^3-44 N_1^2+108 N_1-74\right)+6 \beta  (86 N_1-163)+297\right)\right)\\
   &-\frac{g_s^4}{8g^4\delta^{16}\beta^4}\left(-6048 N_0^6 \beta ^5+16 N_0^5 \beta ^4 (\beta  (4232 N_1+982)-1549)-4 N_0^4 \beta ^3 \left(4 \beta ^2 \left(7088 N_1^2-3492 N_1+1033\right)-2 \beta  (3596
   N_1\right.\right.\\
   &\left.\left.+5973)+10325\right)+8 N_0^3 \beta ^2 \left(\beta ^2 \left(3232 N_1^2-29640 N_1-4326\right)+16 \beta ^3 \left(456 N_1^3-1088 N_1^2+1275 N_1+65\right)+\beta  (18132
   N_1\right.\right.\\
   &\left.\left.+6853)-4350\right)-N_0^2 \beta  \left(4 \beta ^2 \left(27696 N_1^2-29580 N_1+5981\right)+32 \beta ^3 \left(740 N_1^3-7254 N_1^2+6111 N_1-325\right)+16 \beta ^4
   \left(628 N_1^4\right.\right.\right.\\
   &\left.\left.\left.-4216 N_1^3+11172 N_1^2-7232 N_1+105\right)+48 \beta  (131 N_1-581)+14691\right)+N_0 \left(48 \beta ^2 \left(515 N_1^2-1808 N_1-112\right)\right.\right.\\
   &\left.\left.+16 \beta ^3
   \left(1474 N_1^3-7593 N_1^2+9782 N_1+195\right)+32 \beta ^5 N_1 \left(14 N_1^4-279 N_1^3+1522 N_1^2-2956 N_1+1769\right)+8 \beta ^4 \left(488 N_1^4\right.\right.\right.\\
   &\left.\left.\left.-7680 N_1^3+22104 N_1^2-17872
   N_1-105\right)+3 \beta  (7520 N_1+1743)-2448\right)+(2 \beta -1) N_1 \left(6 \beta ^2 \left(632 N_1^2-2905 N_1+2799\right)\right.\right.\\
   &\left.\left.+4 \beta ^3 \left(279 N_1^3-2318 N_1^2+5492
   N_1-3695\right)+16 \beta ^4 \left(7 N_1^4-93 N_1^3+398 N_1^2-658 N_1+353\right)+\beta  (5229 N_1-9795)+2448\right)\right)\\
   &-\frac{g_s^5}{10g^5\delta^{20}\beta^5}\left(93312 N_0^7 \beta ^6-288 N_0^6 \beta ^5 (2 \beta  (2252 N_1+559)-1685)+16 N_0^5 \beta ^4 \left(24 \beta ^2 \left(7711 N_1^2-2423 N_1+1250\right)\right.\right.\\
   &\left.\left.-2 \beta  (31728
   N_1+40813)+67259\right)-40 N_0^4 \beta ^3 \left(4 \beta ^2 \left(1605 N_1^2-47267 N_1-9059\right)+12 \beta ^3 \left(4784 N_1^3-8246 N_1^2+10442 N_1\right.\right.\right.\\
   &\left.\left.\left.+803\right)+2 \beta 
   (62029 N_1+26899)-32463\right)+4 N_0^3 \beta ^2 \left(40 \beta ^2 \left(35489 N_1^2-20156 N_1+10312\right)+40 \beta ^3 \left(5544 N_1^3-66014 N_1^2\right.\right.\right.\\
   &\left.\left.\left.+39818
   N_1-5027\right)+48 \beta ^4 \left(3670 N_1^4-16580 N_1^3+41550 N_1^2-21780 N_1+863\right)-30 \beta  (4710 N_1+14933)+223489\right)\right.\\
   &\left.-2 N_0^2 \beta  \left(60 \beta ^2
   \left(10356 N_1^2-51699 N_1-6938\right)+20 \beta ^3 \left(54900 N_1^3-186600 N_1^2+252646 N_1+13849\right)+40 \beta ^4 \left(4038 N_1^4\right.\right.\right.\\
   &\left.\left.\left.-61004 N_1^3+137454 N_1^2-109848
   N_1-2589\right)+16 \beta ^5 \left(2568 N_1^5-26610 N_1^4+120040 N_1^3-192810 N_1^2+109832 N_1+945\right)+\right.\right.\\
   &\left.\left.\beta  (982802 N_1+373894)-164865\right)+N_0 \left(4 \beta ^2
   \left(246049 N_1^2-415992 N_1+38745\right)+120 \beta ^3 \left(4782 N_1^3-34959 N_1^2+35464 N_1\right.\right.\right.\\
   &\left.\left.\left.-1034\right)+8 \beta ^4 \left(35850 N_1^4-334320 N_1^3+989930 N_1^2-715820
   N_1+7767\right)+64 \beta ^6 N_1 \left(42 N_1^5-1158 N_1^4+9500 N_1^3-31960 N_1^2\right.\right.\right.\\
   &\left.\left.\left.+45679 N_1-22355\right)+16 \beta ^5 \left(2064 N_1^5-47710 N_1^4+240920 N_1^3-460650 N_1^2+265116
   N_1-945\right)+30 \beta  (6880 N_1-4111)\right.\right.\\
   &\left.\left.+50139\right)+(2 \beta -1) N_1 \left(4 \beta ^2 \left(27945 N_1^2-125282 N_1+119563\right)+8 \beta ^3 \left(5895 N_1^3-46950
   N_1^2+108328 N_1-71927\right)\right.\right.\\
   &\left.\left.+8 \beta ^4 \left(1158 N_1^4-14460 N_1^3+59320 N_1^2-95329 N_1+50293\right)+32 \beta ^5 \left(21 N_1^5-386 N_1^4+2480 N_1^3-7080 N_1^2+9025
   N_1\right.\right.\right.\\
   &\left.\left.\left.-4081\right)+6 \beta  (20555 N_1-38242)+50139\right)\right)\\
&+\Ocal(g_s^6)\,,
\end{split}
\end{displaymath}

\begin{displaymath}
\begin{split}
\t\Fcal_D(N_i,\beta)=&\frac{g_s}{4 g \delta^4\beta}\left(-9N_0^3\beta^2+N_1(3+2(-1+N_1)\beta)+2N_0^2\beta(-5+8(1+2N_1)\beta)-N_0(3+2(-5+4N_1)\beta+8(1+N_1+N_1^2)\beta^2)\right)\\
&+\frac{g_s^2}{4g^2\delta^8\beta^2}\left(36N_0^4\beta^3-4N_0^3\beta^2(-19+28(1+2N_1)\beta)+N_1(12+15(-1+N_1)\beta+2(3-5N_1+2N_1^2)\beta^2)\right.\\
&\left.+N_0^2\beta(53+2(-77+12N_1)\beta+4(29+36N_1+36N_1^2)\beta^2)-N_0(-12+(53+88N_1)\beta+2(-39-68N_1+24N_1^2)\beta^2\right.\\
&\left.+8(5+11N_1+3N_1^2+2N_1^3)\beta^3\right)\\
&+\frac{g_s^3}{12g^3\delta^{12}\beta^3}\left(-864 N_0^5 \beta ^4+8 N_0^4 \beta ^3 (466 \beta  (2 N_1+1)-331)-8 N_0^3 \beta ^2 \left(4 \beta ^2 \left(262 N_1^2+262 N_1+189\right)-4 \beta  (29 N_1+260)+383\right)\right.\\
&\left.+4 N_0^2
   \beta  \left(20 \beta ^2 \left(33 N_1^2-174 N_1-109\right)+4 \beta ^3 \left(152 N_1^3+228 N_1^2+622 N_1+273\right)+2 \beta  (965 N_1+781)-393\right)\right.\\
   &\left.-N_0 \left(8 \beta ^2
   \left(419 N_1^2+98 N_1+398\right)+8 \beta ^3 \left(112 N_1^3-546 N_1^2-412 N_1-381\right)+32 \beta ^4 \left(5 N_1^4+10 N_1^3+80 N_1^2+75 N_1+37\right)\right.\right.\\
   &\left.\left.+12 \beta  (88
   N_1-131)+297\right)+N_1 \left(24 \beta ^2 \left(11 N_1^2-27 N_1+16\right)+8 \beta ^3 \left(5 N_1^3-22 N_1^2+32 N_1-15\right)+516 \beta  (N_1-1)+297\right)\right)\\
   &+\frac{g_s^4}{8g^4\delta^{16}\beta^4}\left(6048 N_0^6 \beta ^5-16 N_0^5 \beta ^4 (2116 \beta  (2 N_1+1)-1549)+4 N_0^4 \beta ^3 \left(4 \beta ^2 \left(7088 N_1^2+7088 N_1+4771\right)-2 \beta  (3596
   N_1+13449)\right.\right.\\
   &\left.\left.+10325\right)-8 N_0^3 \beta ^2 \left(\beta ^2 \left(3232 N_1^2-34536 N_1-21990\right)+24 \beta ^3 \left(304 N_1^3+456 N_1^2+1054 N_1+451\right)+3 \beta  (6044
   N_1+5433)\right.\right.\\
   &\left.\left.-4350\right)+N_0^2 \beta  \left(12 \beta ^2 \left(9232 N_1^2+7404 N_1+11455\right)+32 \beta ^3 \left(740 N_1^3-5286 N_1^2-5271 N_1-4010\right)+16 \beta ^4
   \left(628 N_1^4+1256 N_1^3\right.\right.\right.\\
   &\left.\left.\left.+7236 N_1^2+6608 N_1+3085\right)+48 \beta  (131 N_1-1483)+14691\right)-N_0 \left(48 \beta ^2 \left(515 N_1^2-1296 N_1-758\right)+16 \beta ^3
   \left(1474 N_1^3\right.\right.\right.\\
   &\left.\left.\left.-789 N_1^2+5814 N_1+3023\right)+8 \beta ^4 \left(488 N_1^4-3552 N_1^3-2496 N_1^2-9680 N_1-4401\right)+32 \beta ^5 \left(14 N_1^5+35 N_1^4+490 N_1^3\right.\right.\right.\\
   &\left.\left.\left.+700 N_1^2+937
   N_1+353\right)+3 \beta  (7520 N_1+4897)-2448\right)+N_1 \left(48 \beta ^2 \left(79 N_1^2-191 N_1+112\right)+12 \beta ^3 \left(93 N_1^3-398 N_1^2\right.\right.\right.\\
   &\left.\left.\left.+565 N_1-260\right)+8 \beta
   ^4 \left(14 N_1^4-93 N_1^3+234 N_1^2-260 N_1+105\right)+5229 \beta  (N_1-1)+2448\right)\right)\\
   &+\frac{g_s^5}{10g^5\delta^{20}\beta^5}\left(-93312 N_0^7 \beta ^6+288 N_0^6 \beta ^5 (2252 \beta  (2 N_1+1)-1685)-16 N_0^5 \beta ^4 \left(8 \beta ^2 \left(23133 N_1^2+23133 N_1+14833\right)\right.\right.\\
   &\left.\left.-2 \beta  (31728
   N_1+85145)+67259\right)+40 N_0^4 \beta ^3 \left(4 \beta ^2 \left(1605 N_1^2-60559 N_1-38558\right)+4 \beta ^3 \left(14352 N_1^3+21528 N_1^2+\right.\right.\right.\\
   &\left.\left.\left. 44618 N_1+18721\right)+2
   \beta  (62029 N_1+58538)-32463\right)-4 N_0^3 \beta ^2 \left(20 \beta ^2 \left(70978 N_1^2+79294 N_1+96033\right)+80 \beta ^3 \left(2772 N_1^3\right.\right.\right.\\
   &\left.\left.\left.-29003 N_1^2-31778
   N_1-21998\right)+32 \beta ^4 \left(5505 N_1^4+11010 N_1^3+52315 N_1^2+46810 N_1+20967\right)-30 \beta  (4710 N_1+34231)\right.\right.\\
   &\left.\left.+223489\right)+2 N_0^2 \beta  \left(60 \beta ^2
   \left(10356 N_1^2-51393 N_1-36176\right)+20 \beta ^3 \left(54900 N_1^3+22896 N_1^2+252052 N_1+140723\right)\right.\right.\\
   &\left.\left.+40 \beta ^4 \left(4038 N_1^4-38428 N_1^3-45138 N_1^2-110094
   N_1-50559\right)+16 \beta ^5 \left(2568 N_1^5+6420 N_1^4+63600 N_1^3+88980 N_1^2\right.\right.\right.\\
   &\left.\left.\left.+110042 N_1+40405\right)+\beta  (982802 N_1+915556)-164865\right)-N_0 \left(\beta ^2
   \left(984196 N_1^2+99616 N_1+937156\right)+120 \beta ^3 \left(4782 N_1^3\right.\right.\right.\\
   &\left.\left.\left.-19317 N_1^2-10208 N_1-12766\right)+8 \beta ^4 \left(35850 N_1^4-57660 N_1^3+373880 N_1^2+274580
   N_1+194147\right)+16 \beta ^5 \left(2064 N_1^5\right.\right.\right.\\
   &\left.\left.\left.-20630 N_1^4-9820 N_1^3-140460 N_1^2-115344 N_1-58455\right)+64 \beta ^6 \left(42 N_1^6+126 N_1^5+2730 N_1^4+5250 N_1^3+13309
   N_1^2\right.\right.\right.\\
   &\left.\left.\left.+10705 N_1+4081\right)+30 \beta  (6880 N_1-10991)+50139\right)+N_1 \left(540 \beta ^2 \left(207 N_1^2-494 N_1+287\right)+120 \beta ^3 \left(393 N_1^3-1643 N_1^2\right.\right.\right.\\
   &\left.\left.\left.+2284
   N_1-1034\right)+24 \beta ^4 \left(386 N_1^4-2480 N_1^3+6050 N_1^2-6545 N_1+2589\right)+16 \beta ^5 \left(42 N_1^5-386 N_1^4+1450 N_1^3-2750 N_1^2\right.\right.\right.\\
   &\left.\left.\left.+2589 N_1-945\right)+123330
   \beta  (N_1-1)+50139\right)\right)\\
   &+\Ocal(g_s^6)\,.
\end{split}
\end{displaymath}

\end{document}